\documentclass[sigplan,,nonacm]{acmart}
\usepackage{booktabs}
\usepackage{multirow}
\usepackage{array}
\usepackage{subcaption}
\usepackage{enumitem}
\usepackage{xspace}
\usepackage{balance}
\usepackage{tikz}
\usepackage{tabularx}
\usetikzlibrary{arrows.meta,decorations.pathreplacing}
\usepackage[ruled,vlined,linesnumbered]{algorithm2e}

\usepackage{pgfplots}
\pgfplotsset{compat=1.18}

\usepackage{listings}
\definecolor{sysblue}{HTML}{4A90D9}
\definecolor{sysorange}{HTML}{E8833A}
\definecolor{sysgreen}{HTML}{5CB85C}

\newcommand{\sysname}{\textsc{GeoTrain}\xspace}

\newcommand{\paragraphbe}[1]{\noindent\textbf{#1}}

\usepackage[textsize=tiny,disable]{todonotes}

\usepackage{tocloft}

\usepackage{pifont}
\newcommand{\cmark}{\ding{51}}   
\newcommand{\xmark}{\ding{55}}   

\title{\sysname: Energy-Aware Automatic Parallelism for Foundation Model Training on Heterogeneous Geo-Distributed Resources}

\begin{abstract}
Training foundation models at the frontier typically requires large-scale dedicated and centralized clusters with fast interconnects - infrastructure that is prohibitively expensive and generally accessible only to well-resourced organizations. Decentralized training over pooled, geographically distributed GPUs offers an alternative, but introduces substantial heterogeneity in compute, memory, power, and network performance. Existing distributed training systems are primarily designed for clusters with fast, low-latency interconnects and typically use all available GPUs when configuring parallel training. In geo-distributed environments, however, slower devices and weak network links can become bottlenecks, while optimizing solely for throughput overlooks significant energy cost. 
Efficiently selecting and configuring a subset of heterogeneous GPUs therefore requires jointly reasoning about resource characteristics, model partitioning, and pipeline placement.

We present \sysname, an automatic parallelism system that jointly optimizes training performance and energy consumption for heterogeneous geo-distributed GPU environments. \sysname simultaneously selects GPUs, partitions model layers, and determines pipeline configurations while accounting for heterogeneous compute, memory, power, and network constraints. 
To efficiently explore this large combinatorial search space, \sysname employs an optimized memetic algorithm with specialized genetic operators and local search. Experiments with decoder-only Transformer models on real-world and simulated heterogeneous topologies demonstrate that \sysname reduces iteration time and energy consumption by up to $61.4\%$ and $11.7\%$, respectively, over the existing state-of-the-art approaches. It also finds the exact optimum on verifiable instances and scales beyond exact methods as the number of GPUs increases, while exposing favorable energy-time trade-offs, including cases where its time-optimal distributed training plan is both faster and more energy-efficient than the baselines.

\end{abstract}



\author{Bui-Khanh Hoang, TianZhang He, Mohammad Goudarzi}

\authornote{Corresponding author.\newline
  Emails: khanh.hoang@monash.edu, tianzhang.he@monash.edu, mohammad.goudarzi@monash.edu.
}
\affiliation[obeypunctuation=true]{%
  \institution{Faculty of Information Technology, Monash University},
  \city{Melbourne}, \country{Australia}
}

\begin{document}







\maketitle
\begin{figure*}[t]
    \centering
    \includegraphics[width=0.85\textwidth]{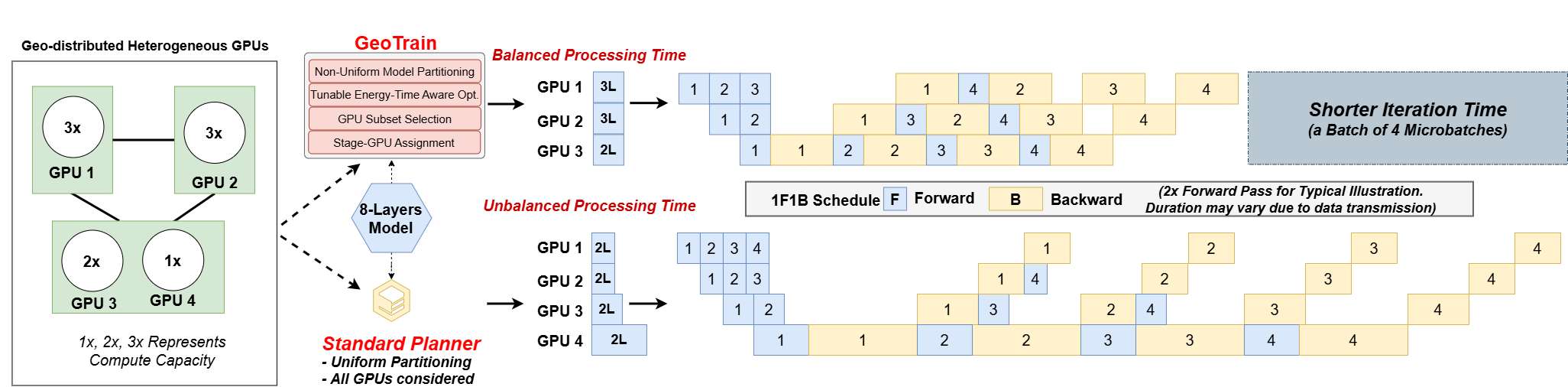}
    \caption{A high-level comparison between \sysname{} and a standard uniform planer using all available GPUs.}
    \label{fig:geotrain-diagram}
\end{figure*}
\section{Introduction}
\label{sec:intro}
Training foundation models, such as GPT-3 and LLaMA, requires distributing computation across hundreds or thousands of GPUs~\cite{brown2020language, touvron2023llama}. As models continue to scale, organisations are increasingly pooling all available GPU resources, including heterogeneous hardware spanning multiple generations and geographically distributed clusters linked by wide-area networks with diverse bandwidth and latency characteristics. Making efficient use of such infrastructure is essential to reducing both training time and the considerable energy expenditure of large-scale model training.

\begin{figure}[t]
  \centering
  \begin{subfigure}[t]{0.5\linewidth}
    \centering
    \includegraphics[width=0.95\linewidth, height=3.5cm]{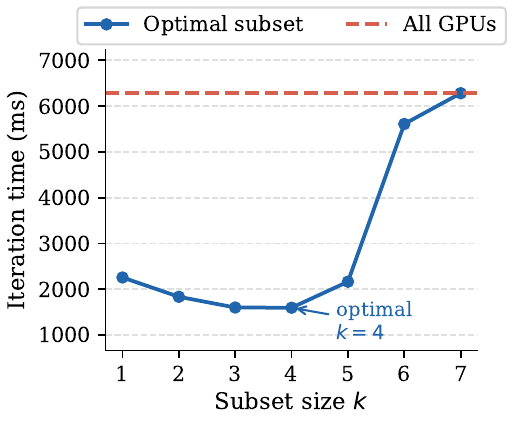}
    \caption{Selecting an optimal GPU subset ($k{=}4$) yields lower
      iteration time than using all GPUs.}
    \label{fig:motivation-subset}
  \end{subfigure}
  \hfill
  \begin{subfigure}[t]{0.46\linewidth}
    \centering
    \includegraphics[width=0.95\linewidth, height=3.2cm]{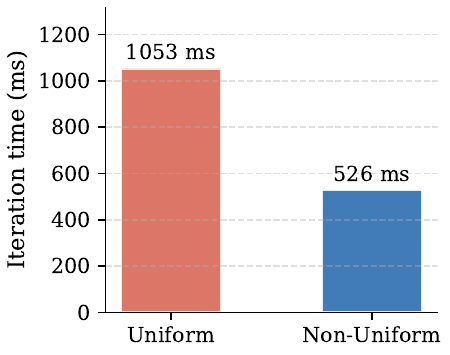}
    \caption{Non-uniform layer assignment eliminates the bottleneck
      stage under uniform one.}
    \label{fig:motivation-stages}
  \end{subfigure}
  \caption{Motivating observations on a real-world heterogeneous geo-distributed configuration with 7 GPUs}
  \label{fig:motivation}
  \vspace{-3mm}
\end{figure}

Systems such as Alpa~\cite{zheng2022alpa} and AMP~\cite{li2022amp} have advanced the automation of parallelism planning for distributed training. Also, a separate line of work, DTFM~\cite{yuan2022dtfm}, SWARM~\cite{ryabinin2023swarm}, Sailor~\cite{strati2025sailor}, Metis~\cite{um2024metis}, and CocktailSGD~\cite{wang2023cocktailsgd}, has begun to address training over heterogeneous or geo-distributed resources directly. However, across these efforts, three major assumptions persist that limit their applicability to the heterogeneous and  geo-distributed paradigm:


\paragraphbe{Full-GPUs participation.}
Existing systems assume full-cluster participation, where all available GPUs are always used for training. This reduces the problem to optimizing configurations over a fixed set of devices, avoiding the combinatorial complexity of GPU subset selection. However, this assumption becomes inefficient in heterogeneous, geo-distributed settings: slow GPUs and weak interconnects can dominate iteration time and negate the benefit of additional parallelism. As shown in Fig.~\ref{fig:motivation-subset}, selecting an optimal subset of $k=4$ GPUs from 7 heterogeneous devices reduces iteration time by 75\% compared to using the full cluster. This demonstrates that carefully chosen subsets can outperform full-cluster execution in heterogeneous environments.


\paragraphbe{Uniform stage partitioning.}
Prior systems partition model layers uniformly across pipeline stages, implicitly assuming homogeneous compute capacity across GPUs. This assumption breaks under heterogeneity, where uniform assignment creates straggler stages on weaker or bandwidth-constrained devices that stall the entire pipeline. Capacity-aware non-uniform partitioning that matches layer allocation to GPU compute and memory constraints yields a more balanced pipeline and significantly lower iteration time (Fig.~\ref{fig:motivation-stages}).

\paragraphbe{Energy-unaware optimization.}
Existing automatic parallelism systems optimize solely for training throughput or iteration time, ignoring energy consumption. However, as the environmental and financial costs of large-scale training continue to rise~\cite{strubell2019energy,patterson2021carbon}, energy efficiency has become a critical objective. Importantly, equivalent training speeds can exhibit substantially different energy profiles across pipeline configurations due to variations in GPU selection, layer placement, and micro-batch sizing, especially in heterogeneous environments with diverse power characteristics.


In this paper, we present \sysname, a system that automatically discovers efficient pipeline parallelism plans for training on heterogeneous, geo-distributed GPUs. To address limitations of prior work, it employs an optimized Memetic Algorithm (MA) with domain-specific operators to co-optimize three coupled decisions: \emph{which} GPUs to use, \emph{how} to partition model layers, and \emph{where} to place each partition. 

\sysname encodes a plan as an integer chromosome, capturing all three decisions in one encoding. A key design is its fitness function: an analytical model that scores each candidate by both iteration time and energy under the heterogeneous, geo-distributed setting, so the entire search is driven toward configurations that jointly minimize iteration time and energy. The MA evolves chromosomes using specialized genetic operators (i.e., crossover, mutation, and local search). The crossover cuts at training stage boundaries so offspring inherit whole stages from each parent, greedily keeps the fittest cut, and occasionally allows a random mid-stage cut for diversity. The mutations employ different strategies at varying granularity to refine solutions gradually, escape local optima, and broaden exploration. A local search then finds the training pipeline's bottleneck and rebalances the workload to relieve it. Figure~\ref{fig:geotrain-diagram} illustrates the resulting scheduling effect by comparing \sysname{} with standard uniform planning using all available GPUs. Together, these components let \sysname explore the plan space efficiently and converge quickly toward balanced, low-cost pipelines.


In designing \sysname, we make following contributions:

\begin{itemize}[leftmargin=*,topsep=2pt,itemsep=1pt]
\item \textbf{Heterogeneity-aware Pipeline Planning}
Unlike prior systems that typically assume all available GPUs must be used, \sysname jointly selects a subset of GPUs and determines a non-uniform layer partition across pipeline stages. It accounts for heterogeneous compute capacities and asymmetric network links to avoid slow devices and poorly connected links, while balancing per-stage cost to eliminate stragglers. This enables \sysname to construct balanced pipelines that achieve lower iteration time than full-cluster or uniform partitioning.

      \item \textbf{Time-Energy-aware optimization.}
        We formulate parallelism planning as an
        optimization that jointly minimizes iteration time and energy
        consumption. An analytical energy cost model captures
        GPU-specific power profiles under the 1F1B pipeline schedule ~\cite{narayanan2019pipedream,
narayanan2021megatron}, while a user-tunable weighting parameter allows operators to navigate the time-energy trade-off. By varying the parameter, \sysname can produce energy-saving, balanced, or time-optimal training configurations, enabling operators to select operating points along the Pareto frontier according to their performance and energy requirements.
  \item \textbf{Domain-specific Memetic Algorithm.}
    We design a domain-specific Memetic Algorithm tailored to the combinatorial planning space of heterogeneous, geo-distributed pipeline training. It combines stage-aware crossover, multi-granularity mutation, and bottleneck-driven local search to efficiently explore candidate configurations and refine promising solutions.
  \item \textbf{Extensive real-world experiments.} We evaluate \sysname{} using both real-world experiments and simulation on heterogeneous geo-distributed GPU configurations with Transformer models examining scalability, optimality, network sensitivity, and energy-time trade-offs. 
\end{itemize}

The remainder of this paper is organized as follows.
We analyze distributed model training (Sect.~\ref{sec:background}) and related work (Sect.~\ref{sec:background:related-work}).
Section~\ref{sec:problem} formalizes the problem.
Section~\ref{sec:design} presents \sysname design.
Section~\ref{sec:eval} analyzes the system and the results.
Section~\ref{sec:conclusion} concludes and discusses future work.

\section{Preliminaries}
\label{sec:background}
Training foundation models involves repeatedly processing training batches through forward and backward passes to compute gradients. When models exceed single-GPU memory, distributed training partitions computation across multiple GPUs using parallelism strategies.

\subsection{Parallelism Methods}
\vspace{-5mm}
\label{sec:background:parallelism}
\mbox{}\par\noindent Distributed training commonly uses three forms of parallelism:
\textbf{Pipeline parallelism (PP)} partitions model layers into stages, each mapped to different devices~\cite{huang2019gpipe,narayanan2019pipedream}. Micro-batches flow through the pipeline, with communication occurring at stage boundaries for activations and gradients.


We focus on PP because its communication is confined to stage boundaries, making it well suited to geo-distributed environments compared with Data parallelism (DP)~\cite{li2020pytorchddp}, which required frequent gradient synchronization, and Tensor parallelism (TP)~\cite{shoeybi2019megatron}, which incurs substantial fine-grained intra-layer communication overhead. The work could be extended to incorporate DP and TP to exploit high-speed interconnects between co-located GPUs, but this is outside our current focus and left for future work.
\subsection{Pipeline Schedules}
\vspace{-5mm}
\label{sec:background:schedules}
\mbox{}\par\noindent Given a pipeline of $Q$ stages processing $M$ micro-batches, the
pipeline schedule determines the ordering of forward and backward
computations across stages. In this work, we adopt the \textbf{1F1B} schedule~\cite{narayanan2019pipedream,
narayanan2021megatron}, which gives the following
iteration time formula:
\begin{equation}
  T^{\text{iter}}(M, u) = \sum_{q=1}^{Q} \tau_q(u)
  + (M-1) \cdot \max_{q=1,\ldots,Q} \tau_q(u)
  \label{eq:1f1b-preview}
\end{equation}
where $\tau_q(u) = \tau_q^F(u) + \tau_q^B(u)$ is the forward-backward service time of stage $q$ for micro-batch size $u$. The first term captures the latency of a micro-batch through the pipeline, including warmup and drain, while the second reflects the steady-state throughput determined by the bottleneck stage.

\subsection{Decoder-Only Transformer Architecture}
\vspace{-3mm}
\label{sec:background:transformer}
\mbox{}\par\noindent We consider a decoder-only Transformer~\cite{vaswani2017attention}, a widely used foundation-model architecture, with hidden size $H$ and $L$ layers. Each layer consists of a multi-head self-attention block with 4 $H\times H$ projections followed by a feed-forward network (FFN) with structure $H \rightarrow 4H \rightarrow H$. The parameter count per layer is:
\begin{equation}
  P_\ell \approx 4H^2 + 8H^2 = 12H^2,
  \label{eq:params-per-layer}
\end{equation}
where the first term covers the attention projections and the second covers the FFN. Biases and LayerNorm parameters are $O(H)$ and omitted.
The matrix multiplication (matmul) cost to process one token through one layer in the
forward pass is approximately~\cite{narayanan2021megatron}:
\begin{equation}
  C_\ell \approx 2 P_\ell \text{ FLOPs/token},
  \label{eq:flops-per-token}
\end{equation}
The backward pass requires roughly $2\times$ the forward FLOPs~\cite{kaplan2020scaling}.
\begin{table}[t]
\centering
\caption{Comparison of \sysname{} with heterogeneous and
geo-distributed foundation-model training systems.}
\label{tab:related-comparison}
\small
\setlength{\tabcolsep}{5pt}
\renewcommand{\arraystretch}{1.25}

\resizebox{\columnwidth}{!}{%
\begin{tabular}{l ccc cc l cc}
\toprule
\textbf{Work}
& \multicolumn{3}{c}{\textbf{Environment}}
& \multicolumn{2}{c}{\textbf{Search Space}}
& \textbf{Method}
& \multicolumn{2}{c}{\textbf{Objective}} \\
\cmidrule(lr){2-4}
\cmidrule(lr){5-6}
\cmidrule(lr){8-9}

& \textbf{Comp.} & \textbf{Net.} & \textbf{Asym.}
& \textbf{Non-unif.} & \textbf{GPU}
&
& \textbf{T} & \textbf{E} \\
& \textbf{hetero.} & \textbf{aware} & \textbf{links}
& \textbf{partition} & \textbf{subset}
&
& & \\
\midrule

\cite{zheng2022alpa}
& \xmark & \cmark & \xmark
& \cmark & \xmark
& ILP+DP
& \cmark & \xmark \\

\cite{li2022amp}
& \cmark & \cmark & \xmark
& \cmark & \xmark
& DP
& \cmark & \xmark \\

\cite{park2020hetpipe}
& \cmark & \xmark & \xmark
& \xmark & \xmark
& Manual
& \cmark & \xmark \\

\cite{ryabinin2023swarm}
& \cmark & \cmark & \xmark
& \xmark & \xmark
& Dynamic
& \cmark & \xmark \\

\cite{yuan2022dtfm}
& \xmark & \cmark & \xmark
& \xmark & \xmark
& EA
& \cmark & \xmark \\

\cite{zong2025hgtrainer}
& \cmark & \cmark & \xmark
& \cmark & \xmark
& MILP+H
& \cmark & \xmark \\

\cite{strati2025sailor}
& \cmark & \cmark & \xmark
& \cmark & \cmark
& DP+H
& \cmark & \xmark \\

\cite{um2024metis}
& \cmark & \xmark & \xmark
& \cmark & \xmark
& DFS
& \cmark & \xmark \\

\midrule

\textbf{Ours}
& \cmark & \cmark & \cmark
& \cmark & \cmark
& MA
& \cmark & \cmark \\

\bottomrule
\end{tabular}%
}

\vspace{2pt}

\parbox{\columnwidth}{%
\footnotesize
\textit{T}: iteration time; \textit{E}: energy.
\textit{ILP}: integer linear programming;
\textit{MILP}: mixed-integer linear programming;
\textit{DP}: dynamic programming;
\textit{EA}: evolutionary algorithm;
\textit{DFS}: depth-first search;
\textit{H}: heuristic pruning;
\textit{MA}: memetic algorithm.
GPU subset: fewer than all available GPUs may be selected.
}

\end{table}
\section{Related Work}
\label{sec:background:related-work}

In this section, we review relevant work, identify its gaps, and position our contributions.

\paragraphbe{Automatic parallelism planning.}
Alpa~\cite{zheng2022alpa} automates inter- and intra-operator parallelism using dynamic programming and integer linear programming, respectively. AMP~\cite{li2022amp} uses a heterogeneity-aware cost model for uneven layers and heterogeneous accelerators, while Galvatron~\cite{miao2022galvatron} automates hybrid parallelism via dynamic programming. These works automate parallelism planning but do not address geo-distributed topology or energy-aware optimization.

\paragraphbe{Heterogeneous and geo-distributed training.}
DTFM~\cite{yuan2022dtfm} addresses heterogeneous networks with explicit bandwidth and latency, but operates on a given parallelization plan. SWARM~\cite{ryabinin2023swarm} targets poorly connected, heterogeneous devices through randomized, adaptively rebalanced pipelines, while CocktailSGD~\cite{wang2023cocktailsgd} reduces communication through gradient compression. Sailor~\cite{strati2025sailor} and HGTrainer~\cite{zong2025hgtrainer} address heterogeneous geo-distributed resources; HetPipe~\cite{park2020hetpipe} combines pipelined model and data parallelism across heterogeneous GPUs. Oobleck~\cite{jang2023oobleck} focuses on fault-tolerant pipeline re-instantiation. Metis~\cite{um2024metis} automatically partitions workloads across heterogeneous GPUs but does not target geo-distributed topology or energy-aware planning.

\paragraphbe{Pipeline parallelism and memory.}
GPipe~\cite{huang2019gpipe} popularizes micro-batch pipelining, while PipeDream~\cite{narayanan2019pipedream} introduces the 1F1B schedule adopted by many later works. GraphPipe~\cite{jeon2025graphpipe} introduces graph pipeline parallelism for general DNN computation graphs by jointly optimizing graph partitioning and micro-batch scheduling. AdaPipe~\cite{sun2024adapipe} jointly optimizes recomputation and stage partitioning for compute-memory balance, rather than network and hardware heterogeneity. ZeRO~\cite{rajbhandari2020zero} and Megatron-LM~\cite{shoeybi2019megatron} target intra-cluster parallelism, complementary to our WAN-pipeline focus.

\paragraphbe{Energy-aware foundation model training.}
Zeus~\cite{you2023zeus} studies GPU power and energy consumption, motivating our energy-aware planning model. Strubell et~al.~\cite{strubell2019energy} and Patterson et~al.~\cite{patterson2021carbon} quantify training's energy and carbon footprint. However, these works do not incorporate energy into automatic parallelism planning.

Across these lines of work, no existing planner combines the capabilities required for heterogeneous, geo-distributed training, as Table~\ref{tab:related-comparison} summarizes. \sysname{} addresses these gaps by supporting directed network topologies with asymmetric links, jointly selecting GPU subsets and non-uniform layer partitions, and incorporating energy into plan search.

\section{System Model and Problem Formulation}
\label{sec:problem}
This section formalizes \sysname{}'s planning problem. We model resource constraints and costs, define the solution space, and formulate the optimization problem.

\subsection{Hardware Model}
\label{sec:problem:hardware}
We model the GPU network as a directed graph
$\mathcal{G} = (G, E)$, where $G = \{g_1, g_2, \ldots, g_N\}$ is the
set of $N$ available GPUs and $E$ is the set of directed
communication links between them. 
It captures three key characteristics of geo-distributed environments often overlooked in prior work: heterogeneous GPU performance, asymmetric WAN bandwidth, and non-uniform latencies ranging from intra-rack microseconds to intercontinental hundreds of milliseconds.
Each GPU $g_i \in G$ is characterized by three properties:

\textbf{Effective compute capacity}:
$\tilde{F}_i = U_i^{\text{comp}} \cdot F_i$, where $F_i$ is the peak
compute capacity (FLOPS) and $U_i^{\text{comp}} \in [0,1]$ is the
fraction effectively available for training.
\textbf{Effective memory}:
$\tilde{M}_i = U_i^{\text{mem}} \cdot M_i$, where $M_i$ is the total
memory capacity (GB) and $U_i^{\text{mem}} \in [0,1]$ is the fraction
usable after framework and runtime overhead.
\textbf{Idle and active power}:
$P_i^{\text{idle}}$ and $P_i^{\text{act}}$ (watts), where
$P_i^{\text{act}}$ is the power drawn when computing at the effective
compute capacity $\tilde{F}_i$.

For each directed edge $(g_i, g_j) \in E$: a communication link
from GPU $i$ to $j$, we define:
\textbf{Effective bandwidth}:
$\tilde{B}_{ij} = \beta_{ij} \cdot B_{ij}$, where $B_{ij}$ is the
nominal link bandwidth (Gbps) and $\beta_{ij} \in [0,1]$ is its
utilization. 
\textbf{One-way latency}:
$L_{ij}$, the time required for data to travel from GPU $i$ to $j$ over the link (ms).

\subsection{Memory Model}
\label{sec:problem:memory}

We consider a decoder-only Transformer. The
training dataset is split into batches of $B$ samples each. Each batch
is further divided into $M$ micro-batches of size $u = B/M$ samples,
where each sample is a sequence of $T$ tokens.

\paragraphbe{Per-layer memory.}
The training-time memory footprint of one Transformer layer consists of parameters with optimizer state plus activations stored for backpropagation:
\begin{equation}
  \text{Mem}_\ell^{\text{train}} \approx
  4\,P_\ell\,\kappa + (Q+1)\,\alpha\,\kappa\,u\,T\,H
  \label{eq:mem-per-layer}
\end{equation}
where $\kappa$ is the bytes per element (4 for FP32, 2 for FP16); $4 P_\ell \kappa$ covers the memory size of parameters, gradients, and the two Adam moment estimates~\cite{kingma2015adam}; $\alpha \approx 10$ is the number of intermediate tensors retained per layer for backpropagation; and $(Q+1)$ bounds the number of micro-batches whose activations the worst-case stage retains during 1F1B warmup. 

\paragraphbe{Per-stage memory.}
For a stage containing $n_q$ consecutive layers, the total memory
footprint is the per-layer cost scaled by the number of layers:
\begin{equation}
  \text{Mem}_q^{\text{train}} \approx
  n_q\,\bigl(4\,P_\ell\,\kappa + (Q+1)\,\alpha\,\kappa\,u\,T\,H\bigr)
  \label{eq:mem-stage}
\end{equation}

\paragraphbe{Last-stage memory.}
The final stage incurs extra terms of language model head
projection ($H \to V$) and its output:
\begin{equation}
  \text{Mem}_{\text{last}}^{\text{train}} \approx
  \text{Mem}_q^{\text{train}}
  + \underbrace{4\,P_{\text{lm}}\,\kappa}_{\text{LM head + Adam}}
  + \underbrace{2\,\kappa\,u\,T\,V}_{\text{logits + gradient}}
  \label{eq:mem-last}
\end{equation}
where $P_{\text{lm}} = H \times V$ is the LM head parameter count and
$V$ is the vocabulary size. Eq.~\ref{eq:mem-stage} and
Eq.~\ref{eq:mem-last} determines the memory
constraint enforced in Eq.~\ref{eq:mem-constraint}.

\subsection{Performance Model}
\label{sec:problem:perf}

We derive the per-stage service time used in the 1F1B iteration-time
formula (Eq.~\ref{eq:1f1b-preview}).

\paragraphbe{Stage computation time.}
Stage $q$ with $n_q$ layers placed on GPU $g_q$ processes one
micro-batch of size $u$ through forward and backward passes in time
\begin{equation}
  \tau_q^{\text{comp}}(u) =
  \frac{(1+\gamma)\,n_q\,C_\ell\,u\,T}{\tilde{F}_{g_q}},
  \label{eq:tau-comp}
\end{equation}
where $\gamma \approx 2$ accounts for the backward pass costing roughly
twice the forward, $C_\ell \approx 2 P_\ell \approx 24 H^2$ is the
per-token forward compute cost of one Transformer layer, and $\tilde{F}_{g_q}$ is the
effective compute capacity of the assigned GPU. The last stage adds
the LM head projection:
\begin{equation}
  \tau_Q^{\text{comp}}(u) =
  \frac{(1+\gamma)\,(n_Q\,C_\ell\,u\,T + 2\,u\,T\,H\,V)}
       {\tilde{F}_{g_Q}}.
  \label{eq:tau-comp-last}
\end{equation}
The last stage computes its $n_Q$ layers like any other stage, but additionally projects the hidden states to vocabulary logits: the extra term $2\,u\,T\,H\,V$ is the cost of projecting each of the $u\,T$ tokens from hidden dimension $H$ to vocabulary size $V$. The factor $(1+\gamma)$
again accounts for the backward pass.

\paragraphbe{Forward communication time.}
Stage $q$ sends activations forward to stage $q+1$ in
\begin{equation}
  \tau_{q \to q+1}^{\text{comm}}(u) =
  L_{g_q, g_{q+1}} + \frac{S_q^{\text{act}}(u)}{\tilde{B}_{g_q, g_{q+1}}},
  \label{eq:tau-fwd}
\end{equation}
where $S_q^{\text{act}}(u) = \kappa \cdot u \cdot T \cdot H$ is the activation tensor size. Transfer time is computed by adding latency to the microbatch's activation size divided by the link effective bandwidth.

\paragraphbe{Backward communication time.}
Gradients of the same size flow back from stage $q$ to stage $q-1$:
\begin{equation}
  \tau_{q \to q-1}^{\text{comm}}(u) =
  L_{g_q, g_{q-1}} + \frac{S_q^{\text{act}}(u)}{\tilde{B}_{g_q, g_{q-1}}}.
  \label{eq:tau-bwd}
\end{equation}
Note that the reverse link $(g_q, g_{q-1})$ may have different
bandwidth and latency than the forward link, which often ignored in prior work but critical in WANs with asymmetric touring.

\paragraphbe{Per-stage service time.}
The total time stage $q$ spends on one micro-batch $u$ is
\begin{equation}
  \tau_q(u) = \tau_q^{\text{comp}}(u)
            + \tau_{q \to q+1}^{\text{comm}}(u)
            + \tau_{q \to q-1}^{\text{comm}}(u),
  \label{eq:tau-q}
\end{equation}
with the first and last stages omitting the incoming and outgoing
communication terms, respectively. Substituting
Eq.~\ref{eq:tau-q} into the 1F1B iteration-time (Eq.~\ref{eq:1f1b-preview}) yields
$T^{\text{iter}}(M, u)$ as a function of the pipeline configuration.

\subsection{Energy Consumption Model}
\label{sec:problem:energy}

We adopt a two-level power model for each GPU $g_q$: idle power
$P_{g_q}^{\text{idle}}$ and active power $P_{g_q}^{\text{act}}$, where
active power is drawn while computing at effective capacity
$\tilde{F}_{g_q}$.

\paragraphbe{Active time per iteration.}
Stage $q$ processes $M$ micro-batches per iteration, so GPU $g_q$
spends
\begin{equation}
  \tau_{g_q}^{\text{active}} \approx M \cdot \tau_q^{\text{comp}}(u)
  \label{eq:tau-active}
\end{equation}
actively computing. The remainder of the iteration,
$T^{\text{iter}} - \tau_{g_q}^{\text{active}}$, is spent idle in
pipeline bubbles or waiting for upstream/downstream stages.

\paragraphbe{Per-GPU energy.}
Over a full iteration, GPU $g_q$ consumes
\begin{equation}
  E_{g_q}^{\text{iter}} \approx
  P_{g_q}^{\text{idle}} \cdot T^{\text{iter}}
  + (P_{g_q}^{\text{act}} - P_{g_q}^{\text{idle}}) \cdot \tau_{g_q}^{\text{active}}.
  \label{eq:energy-per-gpu}
\end{equation}
The first term is the baseline draw over the full iteration; the
second is the incremental cost of active computation.

Total pipeline energy over all $Q$ GPUs in the pipeline is:
\begin{equation}
  E^{\text{iter}} =
  T^{\text{iter}} \cdot \sum_{q=1}^{Q} P_{g_q}^{\text{idle}}
  + M \cdot \sum_{q=1}^{Q} (P_{g_q}^{\text{act}} - P_{g_q}^{\text{idle}})
                          \cdot \tau_q^{\text{comp}}(u).
  \label{eq:energy-total}
\end{equation}

This model exposes a fundamental tension in energy-aware planning. Using fewer, more powerful GPUs reduces idle energy but increases per-GPU power draw, while spreading work across lower-power GPUs reduces peak draw but increases iteration time and idle energy. The optimum depends on workload, hardware, and the desired trade-off.

\subsection{Solution Representation}
\label{sec:problem:solution}
A plan specifies four tightly coupled decisions:
\begin{equation}
  \bigl\{Q,\; \{n_1, \ldots, n_Q\},\; \{g_1, \ldots, g_Q\}\},
  \label{eq:decision-vars}
\end{equation}
where $Q$ is the number of pipeline stages, $1 \leq Q \leq N$; $\{n_1, \ldots, n_Q\}$ are the per-stage layer counts, with $n_q \geq 1$ and $\sum_{q=1}^{Q} n_q = L$, so every stage holds at least one layer and all $L$ layers are covered; and $\{g_1, \ldots, g_Q\}$ is the GPU assignment to stages, constrained to be disjoint, $g_q \neq g_{q'}\;\forall\,q \neq q'$, so no GPU serves more than one stage.



This representation jointly captures GPU subset selection (which $Q$
of $N$ GPUs participate), layer partitioning, GPU-to-stage
assignment. The space grows rapidly with $N$
and $L$, motivating the algorithm in Sect.~\ref{sec:design}.

\subsection{Optimization}
\label{sec:problem:optimization}

GeoTrain combines iteration time and energy into a single scalar objective via a user-tunable weight $\lambda \in [0,1]$:
\begin{equation}
  \min_{Q, \{n_q\}, \{g_q\}, M}
  \;\; \lambda \cdot \frac{T^{\text{iter}}}{T^{\text{ref}}}
     + (1-\lambda) \cdot \frac{E^{\text{iter}}}{E^{\text{ref}}}
  \label{eq:objective}
\end{equation}
subject to
\begin{align}
  \sum_{q=1}^{Q} n_q &= L, \quad n_q \geq 1 \;\;\forall q, \label{eq:layer-constraint} \\
  g_q &\in G, \quad g_q \neq g_{q'} \;\;\forall q \neq q', \label{eq:gpu-constraint} \\
  \text{Mem}_q^{\text{train}} &\leq \tilde{M}_{g_q} \;\;\forall q, \label{eq:mem-constraint} \\
  M &\in \mathbb{Z}^+, \quad 1 \leq M \leq B. \label{eq:mb-constraint}
\end{align}

$T^{\text{ref}}$ and $E^{\text{ref}}$ are normalization constants, typically the time and energy of a baseline plan, chosen so the two
terms are commensurable. At $\lambda = 1$ the problem reduces to pure
time minimization (the regime of prior work). At $\lambda = 0$, it reduces to pure energy minimization. Intermediate values trace the Pareto frontier. The constraints enforce a valid layer partition
(Eq.~\ref{eq:layer-constraint}), distinct per-stage GPU assignment
(Eq.~\ref{eq:gpu-constraint}), per-stage memory feasibility
(Eq.~\ref{eq:mem-constraint}), and a valid micro-batch count
(Eq.~\ref{eq:mb-constraint}).

%

\section{GeoTrain Design}
\label{sec:design}
\sysname{} solves the optimization problem using a Memetic Algorithm (MA)~\cite{moscato2003memetic}. The design comprises a chromosome encoding, domain-specific genetic operators, and a capacity-rebalancing local search.


\subsection{Chromosome Encoding}
\label{sec:design:encoding}
We encode each candidate plan as an integer chromosome of length $L$:
\begin{equation}
  C = [x_1, x_2, \ldots, x_L]
  \label{eq:chromosome}
\end{equation}
where $x_\ell \in \{1, \ldots, N\}$ is the GPU assigned to layer $\ell$.

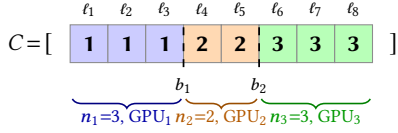
\begin{figure}[t]
  \centering
%

\usetikzlibrary{arrows.meta,decorations.pathreplacing}

\begin{tikzpicture}[
  stbox/.style={draw=black!55, rounded corners=2pt,
                minimum height=0.68cm, font=\scriptsize\sffamily, align=center},
]

\foreach \i in {0,1,2}{
  \fill[blue!20] (\i*0.5,0) rectangle (\i*0.5+0.5,0.5);
  \draw[black!45] (\i*0.5,0) rectangle (\i*0.5+0.5,0.5);
  \node[font=\small\bfseries\sffamily] at (\i*0.5+0.25,0.25) {1};
}
\foreach \i in {3,4}{
  \fill[orange!30] (\i*0.5,0) rectangle (\i*0.5+0.5,0.5);
  \draw[black!45] (\i*0.5,0) rectangle (\i*0.5+0.5,0.5);
  \node[font=\small\bfseries\sffamily] at (\i*0.5+0.25,0.25) {2};
}
\foreach \i in {5,6,7}{
  \fill[green!28] (\i*0.5,0) rectangle (\i*0.5+0.5,0.5);
  \draw[black!45] (\i*0.5,0) rectangle (\i*0.5+0.5,0.5);
  \node[font=\small\bfseries\sffamily] at (\i*0.5+0.25,0.25) {3};
}

\foreach \i/\l in {0/1,1/2,2/3,3/4,4/5,5/6,6/7,7/8}{
  \node[font=\tiny] at (\i*0.5+0.25,0.70) {$\ell_{\l}$};
}

\node[font=\small\sffamily, anchor=east] at (-0.10,0.25) {$C\!=\![$};
\node[font=\small\sffamily, anchor=west] at (4.10,0.25)  {$]$};

\draw[dashed, black!100, line width=0.65pt] (1.5,-0.06) -- (1.5,0.56);
\draw[dashed, black!100, line width=0.65pt] (2.5,-0.06) -- (2.5,0.56);
\node[font=\tiny, black!100, anchor=north] at (1.5,-0.08) {$b_1$};
\node[font=\tiny, black!100, anchor=north] at (2.5,-0.08) {$b_2$};

\draw[decorate, decoration={brace,mirror,amplitude=3.5pt}, blue!50!black, line width=0.6pt]
  (0.03,-0.48) -- (1.47,-0.48);
\node[font=\scriptsize, blue!65!black] at (0.75,-0.72)
  {$n_1{=}3$,\ GPU$_1$};

\draw[decorate, decoration={brace,mirror,amplitude=3.5pt}, orange!60!black, line width=0.6pt]
  (1.53,-0.48) -- (2.47,-0.48);
\node[font=\scriptsize, orange!70!black] at (2.00,-0.72)
  {$n_2{=}2$,\ GPU$_2$};

\draw[decorate, decoration={brace,mirror,amplitude=3.5pt}, green!50!black, line width=0.6pt]
  (2.53,-0.48) -- (3.97,-0.48);
\node[font=\scriptsize, green!60!black] at (3.25,-0.72)
  {$n_3{=}3$,\ GPU$_3$};





  (s2.east) -- (s3.west);

\end{tikzpicture}
  \caption{Chromosome encoding example}
  \label{fig:chromosome}
  \vspace{-3mm}
\end{figure}

The three decision variables in Eq.~\ref{eq:decision-vars} are implicitly encoded. The number of stages $Q$ corresponds to groups of consecutive identical GPU IDs, the layer partition ${n_q}$ to group lengths, and the GPU assignment ${g_q}$ to the GPU ID of each group. Decoding requires a single linear scan. We fix the micro-batch size $u$ and thus the micro-batch count $M=B/u$, excluding $u$ from optimization; jointly optimizing $u$ is left to future work. Figure~\ref{fig:chromosome} shows encoding for $L=8$ and $Q=3$.


A chromosome is \emph{valid} when (i) layers assigned to the same GPU form a contiguous group and (ii) per-stage memory bounds are satisfied (Eq.~\ref{eq:mem-constraint}).



\subsection{GeoTrain Algorithm (Main-Loop) Overview}
\label{sec:design:overview}

Algorithm~\ref{alg:ga} presents the main MA loop. GeoTrain initializes a population and iteratively evolves it using evolutionary operators. The process stops when the runtime limit $T_{\max}$, no-improvement limit $w$, or generation limit $G_{\max}$ is reached.


\paragraphbe{Initial population.}
To seed diversity, GeoTrain samples $P$ chromosomes. For each individual, it uniformly draws a stage count $Q$ from ${1,\ldots,\min(L,N)}$, selects $Q-1$ split points between layers $1$ and $L$ to define stage sizes, samples $Q$ GPUs from $G$, and assigns them to stages in order. Infeasible chromosomes (Eq.~\ref{eq:mem-constraint}) are rejected and resampled. This yields an initial population spanning stage counts and GPU subsets.


\paragraphbe{Selection.}
GeoTrain uses tournament selection~\cite{miller1995tournament}, which provides explicit control over selection pressure via the tournament size, whereas fitness-proportional methods such as roulette-wheel selection are sensitive to the scaling and distribution of fitness values ~\cite{blickle1996comparison}.

\paragraphbe{Survivor replacement.}
Steady-state elitist replacement~\cite{syswerda1991steadystate} is utilized. Each generation preserves the top-$E$ individuals and replaces up to $P - E$ worst incumbents with offspring. Duplicate offspring are discarded to maintain diversity.

\paragraphbe{Fitness function.}
The fitness of a chromosome measures the quality of its encoded pipeline plan. GeoTrain decodes the chromosome to recover the training plan, then computes iteration time $T^{\text{iter}}$ (Eq.~\ref{eq:1f1b-preview}) and per-iteration energy $E^{\text{iter}}$ (Eq.~\ref{eq:energy-total}). These are combined into a single objective via Eq.~\ref{eq:objective}; These are combined into a single objective via Eq.~\ref{eq:objective}; fitness is its negative, so higher fitness indicates a better plan.

\begin{algorithm}[t]
  \small
  \caption{GeoTrain's Memetic Algorithm}
  \label{alg:ga}
  \KwIn{Population size $P$, tournament size $k$, max generations
        $G_{\max}$, runtime limit $T_{\max}$, patience $w$,
        local-search rate $r_{\text{ls}}$, elite size $E$}
  \KwOut{Best plan $C^*$}

  population $\leftarrow$ InitialPopulation($P$) \;
  Evaluate fitness for each $C \in$ population \;
  $C^* \leftarrow \arg\max$ fitness; $g_{\text{stall}} \leftarrow 0$; $t_0 \leftarrow$ current time\;

  \For{generation $\leftarrow 1$ \KwTo $G_{\max}$}{
    \lIf{current time $- t_0 \geq T_{\max}$}{\textbf{break}\tcp*[f]{time limit}}
    \lIf{$g_{\text{stall}} \geq w$}{\textbf{break}\tcp*[f]{no improvement}}

    parents $\leftarrow$ TournamentSelect(population, $k$)\;
    offspring $\leftarrow$ []\;
    \For{iter $\leftarrow 1$ \KwTo $|$parents$|$}{
      $(p_1, p_2) \leftarrow$ random pair from parents\;
      \lIf{$p_1$.genes $= p_2$.genes}{\textbf{skip}}
      offspring.Append(BoundaryCrossover($p_1, p_2$))
      \tcp*[f]{Alg.~\ref{alg:crossover}}
    }
    $C_{\text{best}} \leftarrow$ population.Get\_Best()\;
    \ForEach(\tcp*[f]{Alg.~\ref{alg:mutation}}){$c \in$ offspring}{
      \lIf{$c$.fitness $\leq C_{\text{best}}$.fitness}{
        $c \leftarrow$ Mutate($c$)
      }
    }%
    \ForEach(\tcp*[f]{Alg.~\ref{alg:localsearch}}){$c \in$ offspring}{
      \lIf{rand() $< r_{\text{ls}}$}{
        $c \leftarrow$ LocalSearch($c$)
      }
    }%
    offspring $\leftarrow$ Deduplicate(offspring, population)\;
    population $\leftarrow$ SteadyStateReplace(population, offspring, $E$)\;
    \eIf{best fitness improved}{
      Update $C^*$; $g_{\text{stall}} \leftarrow 0$\;
    }{
      $g_{\text{stall}} \leftarrow g_{\text{stall}} + 1$\;
    }
  }
  \Return{$C^*$}\;
\end{algorithm}

\subsection{Stage-Boundary-Aware Crossover}
\label{sec:design:crossover}
\begin{figure}[t]
  \centering
  \includegraphics[width=\columnwidth]{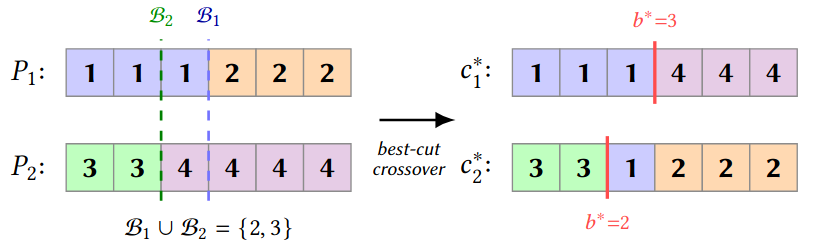}
  \caption{Stage-boundary-aware best-cut crossover ($L=6$). Candidate
    cut positions $\mathcal{B}_1 \cup \mathcal{B}_2 = \{2,3\}$ (dashed
    lines) are drawn from stage boundaries of $P_1$ and $P_2$. For each
    direction, all candidate cuts are evaluated; the cut $b^*$ yielding
    the highest offspring fitness is selected (red tick). $c_1^*$ takes
    $P_1$'s prefix and $P_2$'s suffix; $c_2^*$ takes $P_2$'s prefix and
    $P_1$'s suffix.}
    \vspace{-1.5mm}
  \label{fig:crossover}
\end{figure}

Standard single-point crossover that randomly cuts a chromosome may disrupt the intrinsic stage structure of parent plans. We therefore propose a \emph{stage-boundary-aware best-cut crossover} that restricts candidate cuts to stage boundaries and selects the fitness-maximizing cut (Alg.~\ref{alg:crossover}). Figure~\ref{fig:crossover} illustrates the procedure on a six-layer example. Candidate cuts come from the union of stage boundaries across both parents, $\mathcal{B}_1 \cup \mathcal{B}_2$. Restricting cuts to stage boundaries preserves contiguous, well-balanced stages. Choosing the cut by fitness further steers recombination toward the most beneficial combination. 
Additionally, with probability $p_{\text{fall}}$, we evaluate all possible cuts to explore potentially better offspring beyond stage boundaries (Alg.~\ref{alg:crossover}, line 3).

\begin{algorithm}[t]
  \small
  \caption{Stage-Boundary-Aware Best-Cut Crossover BoundaryCrossover($p_1, p_2$)}
  \label{alg:crossover}
  \KwIn{Parents $P_1, P_2$: $P_1$.genes $\neq P_2$.genes; Fallback probability $p_{\text{fall}}$}
  \KwOut{Offspring list}

  $\mathcal{B}_1, \mathcal{B}_2 \leftarrow$ stage boundaries of $P_1, P_2$\;
  \eIf{rand() $< p_{\text{fall}}$}{
    candidates $\leftarrow \{1, \ldots, L-1\}$ \tcp*{full-scan fallback}
  }{
    candidates $\leftarrow \mathcal{B}_1 \cup \mathcal{B}_2$ \tcp*{boundary-restricted}
  }
  $b_1^* \leftarrow \text{None}$; $f_1^* \leftarrow -\infty$\;
  $b_2^* \leftarrow \text{None}$; $f_2^* \leftarrow -\infty$\;
  \ForEach{$b \in$ candidates}{
    $c_1 \leftarrow P_1[{:}b] \oplus P_2[b{:}]$\tcp*{$\oplus$: concatenation}
    $c_2 \leftarrow P_2[{:}b] \oplus P_1[b{:}]$\;
    \If{$c_1$.Validate() \textbf{and} Fitness$(c_1) > f_1^*$}{
      $b_1^* \leftarrow b$;\quad $f_1^* \leftarrow$ fitness$(c_1)$\;
    }
    \If{$c_2$.Validate() \textbf{and} Fitness$(c_2) > f_2^*$}{
      $b_2^* \leftarrow b$;\quad $f_2^* \leftarrow$ fitness$(c_2)$\;
    }
  }
  offspring $\leftarrow$ []\;
  \lIf{$b_1^* \neq \text{None}$}{offspring.Add($P_1[{:}b_1^*] \oplus P_2[b_1^*{:}]$)}
  \lIf{$b_2^* \neq \text{None}$}{offspring.Add($P_2[{:}b_2^*] \oplus P_1[b_2^*{:}]$)}
  \Return offspring\;
\end{algorithm}

\subsection{Contiguity-Preserving Mutation}
\label{sec:design:mutation}
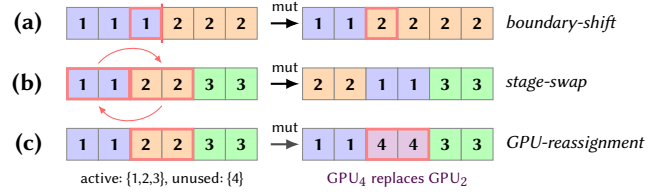
\begin{figure}[t]
  \centering
%
%

\usetikzlibrary{arrows.meta}

\begin{tikzpicture}



\foreach \i/\c in {0/blue!20,1/blue!20,2/blue!20,3/orange!30,4/orange!30,5/orange!30}{
  \fill[\c] (\i*0.42,0) rectangle (\i*0.42+0.42,0.42);
  \draw[black!45] (\i*0.42,0) rectangle (\i*0.42+0.42,0.42);
}
\foreach \i/\g in {0/1,1/1,2/1,3/2,4/2,5/2}{
  \node[font=\scriptsize\bfseries\sffamily] at (\i*0.42+0.21,0.21) {\g};
}
\draw[red!65, line width=0.8pt] (1.26,-0.05) -- (1.26,0.47);
\draw[red!50, line width=1.0pt] (0.84,0.01) rectangle (1.25,0.41);  

\draw[-{Latex[length=5pt,width=4pt]}, thick, black!100]
  (2.70,0.21) -- (3.08,0.21);
\node[font=\tiny\sffamily, above=1pt] at (2.89,0.21) {mut};

\foreach \i/\c in {0/blue!20,1/blue!20,2/orange!30,3/orange!30,4/orange!30,5/orange!30}{
  \fill[\c] (\i*0.42+3.12,0) rectangle (\i*0.42+3.12+0.42,0.42);
  \draw[black!45] (\i*0.42+3.12,0) rectangle (\i*0.42+3.12+0.42,0.42);
}
\foreach \i/\g in {0/1,1/1,2/2,3/2,4/2,5/2}{
  \node[font=\scriptsize\bfseries\sffamily] at (\i*0.42+3.12+0.21,0.21) {\g};
}
\draw[red!50, line width=1.0pt] (3.12+0.84,0.01) rectangle (3.12+1.25,0.41);

\node[font=\small\sffamily, anchor=east] at (-0.20,0.21)
  {\textbf{(a)}};
\node[font=\scriptsize\sffamily, anchor=west] at (5.68,0.21)
  {\textit{boundary-shift}};

\def\dy{-0.8}

\foreach \i/\c in {0/blue!20,1/blue!20,2/orange!30,3/orange!30,4/green!28,5/green!28}{
  \fill[\c] (\i*0.42,\dy) rectangle (\i*0.42+0.42,\dy+0.42);
  \draw[black!45] (\i*0.42,\dy) rectangle (\i*0.42+0.42,\dy+0.42);
}
\foreach \i/\g in {0/1,1/1,2/2,3/2,4/3,5/3}{
  \node[font=\scriptsize\bfseries\sffamily] at (\i*0.42+0.21,\dy+0.21) {\g};
}
\draw[red!50, line width=1.0pt] (0.01,\dy+0.01) rectangle (0.83,\dy+0.41);   
\draw[red!50, line width=1.0pt] (0.85,\dy+0.01) rectangle (1.67,\dy+0.41);   
\draw[-{Latex[length=3pt,width=2.5pt]}, red!60, bend left=50]
  (0.42,\dy+0.48) to (1.26,\dy+0.48);
\draw[-{Latex[length=3pt,width=2.5pt]}, red!60, bend left=50]
  (1.26,\dy-0.08) to (0.42,\dy-0.08);

\draw[-{Latex[length=5pt,width=4pt]}, thick, black!100]
  (2.70,\dy+0.21) -- (3.08,\dy+0.21);
\node[font=\tiny\sffamily, above=1pt] at (2.89,\dy+0.21) {mut};

\foreach \i/\c in {0/orange!30,1/orange!30,2/blue!20,3/blue!20,4/green!28,5/green!28}{
  \fill[\c] (\i*0.42+3.12,\dy) rectangle (\i*0.42+3.12+0.42,\dy+0.42);
  \draw[black!45] (\i*0.42+3.12,\dy) rectangle (\i*0.42+3.12+0.42,\dy+0.42);
}
\foreach \i/\g in {0/2,1/2,2/1,3/1,4/3,5/3}{
  \node[font=\scriptsize\bfseries\sffamily] at (\i*0.42+3.12+0.21,\dy+0.21) {\g};
}

\node[font=\small\sffamily, anchor=east] at (-0.20,\dy+0.21)
  {\textbf{(b)}};
\node[font=\scriptsize\sffamily, anchor=west] at (5.68,\dy+0.21)
  {\textit{stage-swap}};

\def\dy{-1.6}

\foreach \i/\c in {0/blue!20,1/blue!20,2/orange!30,3/orange!30,4/green!28,5/green!28}{
  \fill[\c] (\i*0.42,\dy) rectangle (\i*0.42+0.42,\dy+0.42);
  \draw[black!45] (\i*0.42,\dy) rectangle (\i*0.42+0.42,\dy+0.42);
}
\foreach \i/\g in {0/1,1/1,2/2,3/2,4/3,5/3}{
  \node[font=\scriptsize\bfseries\sffamily] at (\i*0.42+0.21,\dy+0.21) {\g};
}
\draw[red!50, line width=1.0pt] (0.85,\dy+0.01) rectangle (1.67,\dy+0.41);

\node[font=\tiny\sffamily, black!100] at (1.26,\dy-0.25)
  {active: \{1,2,3\}, unused: \{4\}};

\draw[-{Latex[length=5pt,width=4pt]}, thick, black!70]
  (2.70,\dy+0.21) -- (3.08,\dy+0.21);
\node[font=\tiny\sffamily, above=1pt] at (2.89,\dy+0.21) {mut};

\foreach \i/\c in {0/blue!20,1/blue!20,2/violet!22,3/violet!22,4/green!28,5/green!28}{
  \fill[\c] (\i*0.42+3.12,\dy) rectangle (\i*0.42+3.12+0.42,\dy+0.42);
  \draw[black!45] (\i*0.42+3.12,\dy) rectangle (\i*0.42+3.12+0.42,\dy+0.42);
}
\foreach \i/\g in {0/1,1/1,2/4,3/4,4/3,5/3}{
  \node[font=\scriptsize\bfseries\sffamily] at (\i*0.42+3.12+0.21,\dy+0.21) {\g};
}
\draw[red!50, line width=1.0pt] (3.12+0.85,\dy+0.01) rectangle (3.12+1.67,\dy+0.41);
\node[font=\tiny\sffamily, violet!60!black] at (3.12+1.26,\dy-0.25)
  {GPU$_4$ replaces GPU$_2$};

\node[font=\small\sffamily, anchor=east] at (-0.20,\dy+0.21)
  {\textbf{(c)}};
\node[font=\scriptsize\sffamily, anchor=west] at (5.68,\dy+0.21)
  {\textit{GPU-reassignment}};

\end{tikzpicture}
  \vspace{-15pt}
  \caption{Example of three contiguity-preserving mutation operators applied to a six-layer chromosome.}
    \vspace{-3mm}
  \label{fig:mutation}
\end{figure}

\sysname{} applies one of three mutation operators, selected uniformly at random, to each eligible offspring (Alg.~\ref{alg:mutation}; Fig.~\ref{fig:mutation}). All preserve layer contiguity by construction. Mutation targets offspring whose fitness does not exceed the population best, preserving elite solutions while diversifying others. If mutation violates the memory constraint (Eq.~\ref{eq:mem-constraint}), the chromosome reverts to its parent.


\paragraphbe{Boundary-shift mutation.}
Select a random stage boundary position $b$ (an index where
$x_b \neq x_{b-1}$) and set $x_{b-1} \leftarrow x_b$, transferring the last layer of the left stage to the right stage (Alg.~\ref{alg:mutation}, lines 2-3). As the finest-grained move, it can produce a mutant that rebalances load between two adjacent stages, easing the slower one when it bottlenecks the pipeline.

\paragraphbe{Stage-swap mutation.}
Select two pipeline stages uniformly at random and swap their positions (Alg.~\ref{alg:mutation}, lines 7-8). This reorders GPUs across pipeline positions, allowing heavy inter-stage boundaries to move onto faster links and alleviating communication bottlenecks on the asymmetric network, without changing the active GPU subset.

\paragraphbe{GPU-reassignment mutation.}
Reassign a randomly selected stage $q$ to a uniformly sampled inactive GPU $g'\in G\setminus{g_1,\ldots,g_Q}$ (Alg.~\ref{alg:mutation}, lines 10-14). This introduces a new GPU into the active subset while retiring $g_q$, providing the MA's primary mechanism for exploring alternative GPU subsets. If all $N$ GPUs are active, the operator is non-operational.


\begin{algorithm}[t]
  \small
  \caption{Contiguity-Preserving Mutation}
  \label{alg:mutation}
  \KwIn{Copied Offspring $c'$: copy of $c$\;}
  \KwOut{Mutated offspring $c'$}

  Sample mutation type uniformly from \{boundary-shift, stage-swap,
  GPU-reassignment\}\;
  \uIf{boundary-shift}{
    $\mathcal{B} \leftarrow \{i : c'\text{.genes}[i] \neq c'\text{.genes}[i-1],\; i \in 1..L-1\}$\;
    $b \leftarrow$ uniform sample from $\mathcal{B}$\;
    $c'$.genes$[b-1] \leftarrow c'$.genes$[b]$\;
  }
  \uElseIf{stage-swap}{
    $(s_1, s_2) \leftarrow$ TwoStagesSampled($c'$)\;
    $c'$.genes$[s_1.\text{start} : s_2.\text{end}] \leftarrow c'$.genes$[s_2] \oplus c'$.genes$[s_1.\text{end} : s_2.\text{start}] \oplus c'$.genes$[s_1]$\;
  }
  \Else{
    $q \leftarrow$ random stage in $c'$\;
    unused $\leftarrow G \setminus \{$all GPUs in $c'$.genes$\}$\;
    \lIf{unused $= \emptyset$}{\Return $c$}
    $g' \leftarrow$ uniform sample from unused\;
    Reassign all layers in stage $q$ to $g'$\;
  }
  \lIf{$c'$.Validate()}{\Return $c'$}
  \Return $c$\;
\end{algorithm}

\subsection{Capacity-Rebalancing Local Search}
\label{sec:design:localsearch}
The local search exploits the 1F1B cost model to refine offspring toward locally optimal load balance, making \sysname{} memetic rather than purely evolutionary.

\begin{figure}[t]
  \centering

\begin{tikzpicture}


\node[font=\small\bfseries\sffamily] at (1.5,1.15) {Before};

\foreach \i/\c in {0/blue!20,1/blue!20,2/blue!20,3/blue!20,4/orange!30,5/orange!30}{
  \fill[\c] (\i*0.55,0) rectangle (\i*0.55+0.55,0.5);
  \draw[black!45] (\i*0.55,0) rectangle (\i*0.55+0.55,0.5);
}
\foreach \i/\g in {0/1,1/1,2/1,3/1,4/2,5/2}{
  \node[font=\small\bfseries\sffamily] at (\i*0.55+0.275,0.25) {\g};
}

\draw[dashed, gray!65, line width=0.65pt] (2.2,-0.08) -- (2.2,0.58);

\fill[blue!30] (0.0,-1.6) rectangle (2.2,-0.7);
\draw[black!45] (0.0,-1.6) rectangle (2.2,-0.7);
\node[font=\tiny\sffamily] at (1.1,-1.15) {$\tau_1 = 4/\tilde{F}_1$};

\fill[orange!40] (2.2,-1.0) rectangle (3.3,-0.7);
\draw[black!45] (2.2,-1.0) rectangle (3.3,-0.7);
\node[font=\tiny\sffamily] at (2.75,-0.85) {$\tau_2 = 1/\tilde{F}_1$};

\node[font=\tiny\sffamily, red!65!black, anchor=north] at (1.1,-1.68)
  {$\longleftarrow$ bottleneck stage};

\node[font=\scriptsize\sffamily, blue!65!black, anchor=north] at (1,-1.95)
  {GPU$_1$ ($\tilde{F}_1$, slow)};
\node[font=\scriptsize\sffamily, orange!70!black, anchor=north] at (2.75,-1.95)
  {GPU$_2$ ($2\tilde{F}_1$, fast)};


\draw[-{Latex[length=6pt,width=5pt]}, thick, black!100]
  (3.55,-0.55) -- (4.25,-0.55);
\node[font=\tiny\sffamily, align=center, black!100] at (3.9,-0.2)
  {local search\\[-1pt]$\Delta = 2$};


\def\ox{4.5}

\node[font=\small\bfseries\sffamily] at (\ox+1.65,1.15) {After};

\foreach \i/\c in {0/blue!20,1/blue!20,2/orange!30,3/orange!30,4/orange!30,5/orange!30}{
  \fill[\c] (\i*0.55+\ox,0) rectangle (\i*0.55+\ox+0.55,0.5);
  \draw[black!45] (\i*0.55+\ox,0) rectangle (\i*0.55+\ox+0.55,0.5);
}
\foreach \i/\g in {0/1,1/1,2/2,3/2,4/2,5/2}{
  \node[font=\small\bfseries\sffamily] at (\i*0.55+\ox+0.275,0.25) {\g};
}

\draw[dashed, gray!65, line width=0.65pt] (\ox+1.1,-0.08) -- (\ox+1.1,0.58);

\fill[blue!30] (\ox,-1.0) rectangle (\ox+1.1,-0.7);
\draw[black!45] (\ox,-1.0) rectangle (\ox+1.1,-0.7);
\node[font=\tiny\sffamily] at (\ox+0.55,-0.85) {$\tau_1 = 2/\tilde{F}_1$};

\fill[orange!40] (\ox+1.1,-1.0) rectangle (\ox+3.3,-0.7);
\draw[black!45] (\ox+1.1,-1.0) rectangle (\ox+3.3,-0.7);
\node[font=\tiny\sffamily] at (\ox+2.2,-0.85) {$\tau_2 = 2/\tilde{F}_1$};

\node[font=\tiny\sffamily, green!50!black, anchor=north] at (\ox+1.65,-1.08)
  {$\longleftarrow$ load balanced $\longrightarrow$};

\node[font=\scriptsize\sffamily, blue!65!black, anchor=north] at (\ox+0.55,-1.32)
  {GPU$_1$};
\node[font=\scriptsize\sffamily, orange!70!black, anchor=north] at (\ox+2.2,-1.32)
  {GPU$_2$};

\draw[decorate, decoration={brace, amplitude=4pt},
  red!60, line width=0.65pt]
  (1.1,0.52) -- (2.2,0.52);
\node[font=\tiny, red!60!black, anchor=south] at (1.92,0.6)
  {$\Delta = 2$ layers transferred};
\draw[-{Latex[length=3pt]}, red!55, dashed]
  (1.92,0.9) to[bend left=20] (\ox+1.9,0.7);

\end{tikzpicture}
  \caption{Capacity-rebalancing local search example ($L=6$, GPU$_2$ is
    $2\times$ faster than GPU$_1$).}
    \vspace{-3mm}
  \label{fig:localsearch}
\end{figure}
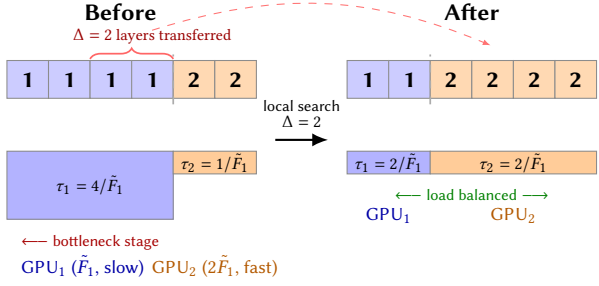

Figure~\ref{fig:localsearch} illustrates the effect on a concrete
two-stage example. The search iteratively applies the convergence step
in Alg.~\ref{alg:localsearch}.
At each step, the assigned GPUs are
ranked by $n_q / \tilde{F}_{g_q}$ (which is proportional to each stage's compute time) in descending order, exposing the current pipeline bottleneck ( Alg.~\ref{alg:localsearch}, lines 2-5). The search identifies the \emph{worst} stage (highest per-layer compute time with available memory headroom) and the \emph{best} stage (lowest per-layer compute time with available memory headroom). If the best stage is strictly faster than the worst, meaning that transferring layers from worst to best would reduce the bottleneck, the search computes the optimal transfer count $\Delta$ (Alg.~\ref{alg:localsearch}, lines 8-10): the maximum number of layers that can be transferred without making the original bottleneck stage faster than the least-loaded stage, thus maintaining a balanced pipeline. The chromosome is then rebuilt with the updated layer counts. The step repeats until \textit{unchanged}, indicating convergence to a local optimum of the load-balancing objective.

\paragraphbe{Interaction with the evolutionary operators.}
Following mutation, local search is applied to offspring with probability $r_{\text{ls}}=0.5$ to balance exploitation and exploration. It refines selected offspring toward balanced load assignments while leaving others unchanged for evolutionary exploration.

\begin{algorithm}[t]
  \small
  \caption{Capacity-Rebalancing Local Search (one convergence step)}
  \label{alg:localsearch}
  \KwIn{Copy of Chromosome $c' \leftarrow$ copy of $c$\;}
  \KwOut{Refined chromosome $c'$}
  \Repeat{$c'$.genes unchanged}{
    Compute load $\ell_q = n_q / \tilde{F}_{g_q}$ for each stage $q$\;
    Sort stages descending by $\ell_q$\;
    worst $\leftarrow$ highest-load stage with transferable layers\;
    best $\leftarrow$ lowest-load stage with available capacity\;
    \lIf{worst $=$ None \textbf{or} best $=$ None \textbf{or}
         $\tilde{F}_{\text{best}} \leq \tilde{F}_{\text{worst}}$}{\textbf{break}}
    $\Delta \leftarrow \min\!\bigl(\text{cap}(\text{best}),\; n_{\text{worst}}\bigr)$\;
    \For{$i \leftarrow 1$ \KwTo $\Delta - 1$}{
      \lIf{$(n_{\text{best}} + i)/\tilde{F}_{\text{best}} \geq
           (n_{\text{worst}} - i)/\tilde{F}_{\text{worst}}$}{$\Delta \leftarrow i$\; \textbf{break}}
    }
    $n_{\text{best}} \leftarrow n_{\text{best}} + \Delta$;\quad
    $n_{\text{worst}} \leftarrow n_{\text{worst}} - \Delta$\;
    Rebuild $c'$.genes from updated layer counts\;
  }
  \Return $c'$\;
\end{algorithm}

\section{Performance Evaluation}
\label{sec:eval}
We evaluate \sysname{} via real-hardware experiments and larger-scale simulation to address these research questions:

\textbf{RQ1.} Can \sysname{} find lower-cost pipeline
        configurations than state-of-the-art baseline on
        heterogeneous GPU environments?
\textbf{RQ2.} How does \sysname{} scale as the number of
        GPUs and model layers grows?
\textbf{RQ3.} How close are \sysname{}'s solutions to the
        globally optimal configuration, and for what environments is
        optimality recoverable?
\textbf{RQ4.} How sensitive is \sysname{} under varying geo-distributed network conditions?
 \textbf{RQ5.} What trade-offs emerge between iteration time
        and energy consumption?

\subsection{Experimental Setup}
\label{sec:eval:setup}
We implement \sysname{} in Python using the 1F1B iteration-time and two-level energy models (Sect.~\ref{sec:problem:energy}). We validate it on real hardware (Sect.~\ref{sec:eval:hardware}), while other experiments use simulated heterogeneous GPU topologies to explore configurations beyond resource limits.

\paragraphbe{Cluster environments.}
Real-hardware experiments run on a six-node cluster, each node equipped with a single GPU: an A100 (40GB), an A30 (24GB), three A40s (48GB each), and T4 (16GB). Nodes are provisioned with 4-60 vCPUs and 16-240GB of host RAM. They are geographically located across different clouds (Nectar\footnote{https://ardc.edu.au/services/ardc-nectar-research-cloud/} and Google Cloud Platform) and connected via a WireGuard VPN exhibiting heterogeneous communication characteristics with bandwidths ranging from 100\,Mbps to 2\,Gbps and latencies from sub-millisecond to 50\,ms. By utilizing virtualization and shared tenancy on the cloud (e.g., NVIDIA's Multi-Instance GPU technology), each GPU slices its full capacity to provide a certain effective FLOPS, which we have to profile before the planner runs.

Regarding simulation, we use synthetic GPU topologies with various compute and memory capacity, and asymmetric network links,  specific details are given per experiment.

\paragraphbe{Model.}
We use a decoder-only Transformer in both real training and simulated studies. Specific configurations are summarized in  Table~A.\ref{tab:configs}. In each case, the
microbatch size $u$ and count $M$ satisfy $u \cdot M = B$ and are fixed per
experiment.

\paragraphbe{Search parameters.} \sysname{} uses population size $P\!=\!200$, tournament size $k\!=\!4$, elite size $E\!=\!2$, local-search rate $r_{\text{ls}}\!=\!0.5$, runtime limit $T_{\max}\!=\!120,\mathrm{s}$, no-improvement limit $w\!=\!100$ generations, and maximum generations $G_{\max}\!=\!500$.


\subsection{Baselines}
We compare against the following baselines:

\paragraphbe{Metis.} A deterministic search baseline adapted from state-of-the-art Metis~\cite{um2024metis} to a PP-only 
, where each stage is restricted to a single GPU. 
The default (dlft.) variant always uses all GPUs, following the original method. The improved variant (impr.) exhaustively searches GPU subsets up to 7 GPUs and uses random sampling beyond. The dlft. is applied in real-hardware experiments and impr. for large-scale simulation, where the original Metis becomes infeasible. 
Both variants use the same analytical cost model as \sysname{}, with the energy disabled, since Metis does not optimize for energy.

\paragraphbe{Brute-Force (BF) Search.}
Exhaustively enumerates pipeline configurations, including GPU subsets, orderings, and layer partitions, yielding the global optimum under the cost model and constraints. It is only practical for small setups.

\paragraphbe{Branch-and-Bound (B\&B).}
Incrementally constructs pipeline configurations and prunes partial solutions using iteration-time and feasibility bounds, preserving optimality while reducing enumeration cost. It is remains practical only for small setups.


\begin{table}[t]
  \centering
  \caption{Real-hardware scaling configurations. Avg. iteration times are measured over iterations 1-9, excluding warm-up (iteration 0).~$\pm$ denotes the coefficient of variation.}
  \label{tab:hardware-plans}
  \small
  \resizebox{0.85\columnwidth}{!}{%
  \setlength{\tabcolsep}{5pt}
    \resizebox{0.7\columnwidth}{!}{
  \begin{tabular}{c c r r r c}
    \toprule
    \textbf{Pool} & \textbf{L}
      & \textbf{\sysname{} (s)} & \textbf{dlft. Metis (s)}
      & \textbf{Gain} & \textbf{GPUs used} \\
    \midrule
    2 & 16 & $27.8\;(\pm0.7\%)$ & $43.5\;(\pm2.2\%)$ & $+$36.3\% & 1 of 2 \\
    4 & 24 & $16.0\;(\pm0.6\%)$ & $41.4\;(\pm0.1\%)$ & $+$61.4\% & 3 of 4 \\
    6 & 32 & $16.7\;(\pm0.9\%)$ & $33.6\;(\pm1.9\%)$ & $+$50.1\% & 5 of 6 \\
    \bottomrule
  \end{tabular}}
  }
\vspace{-2mm}
\end{table}

\subsection{Real-Hardware Evaluation}
\label{sec:eval:hardware}
We evaluate \textbf{RQ1} and \textbf{RQ2} across three real-hardware configurations (Table~\ref{tab:hardware-plans}), varying GPU availability and model depth while holding other factors fixed. Across all configurations, \sysname{} consistently outperforms dlft. Metis, reducing iteration time by 36.3\%-61.4\%. This gain stems from selectively excluding slower devices to mitigate stragglers. As shown in the \textbf{\textit{GPUs used}} column, optimal plans use only a subset of GPUs, avoiding slowdowns from heterogeneous performance and communication. In contrast, dlft. Metis uses all available devices, making it more sensitive to stragglers.

\begin{figure}[t]
  \centering
  \includegraphics[width=0.75\linewidth]{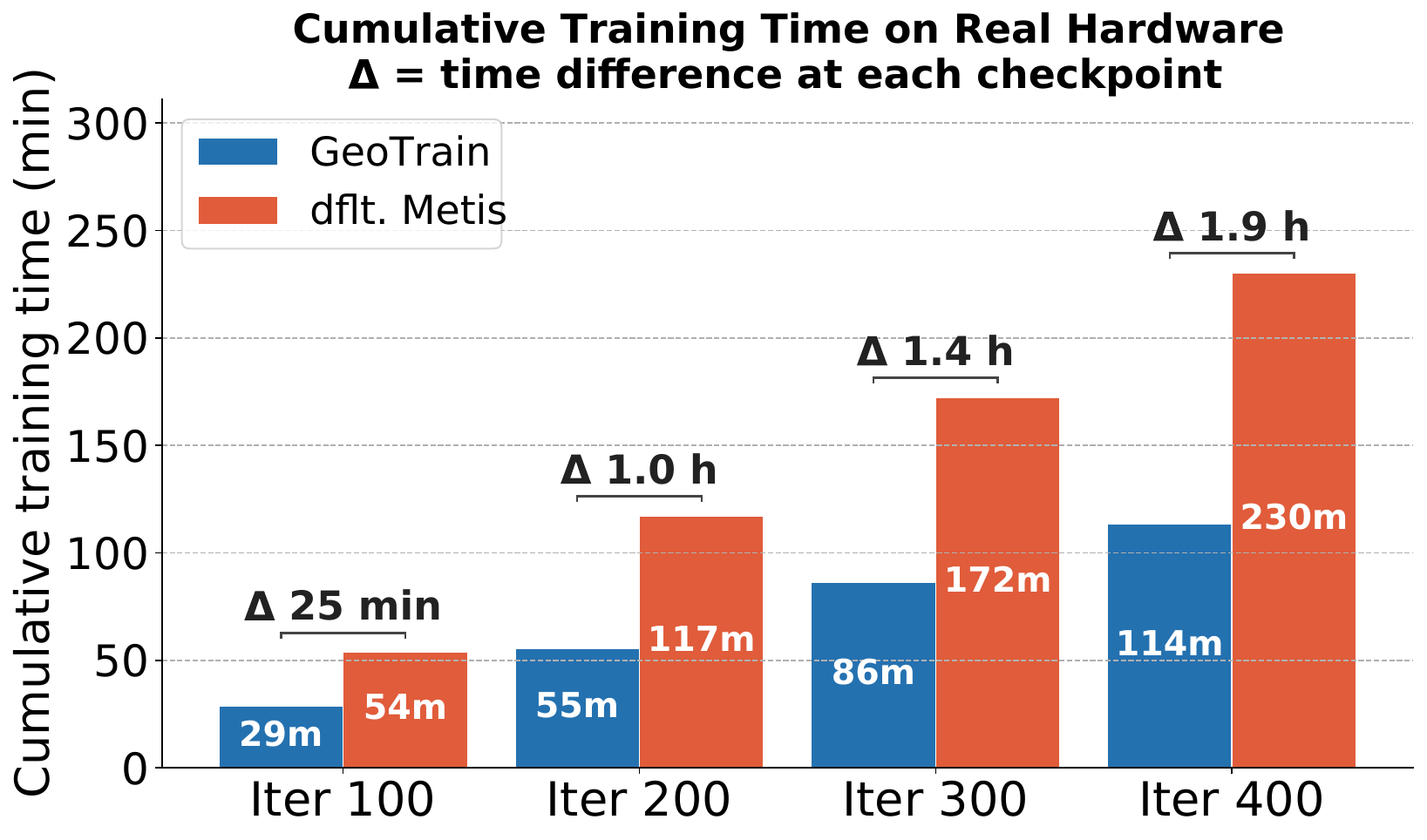}
  \caption{Cumulative training time on real 6-GPU pool}
  \vspace{-3mm}
  \label{fig:cummulative-cost}
\end{figure}

Additionally, the low coefficients of variation in Table~\ref{tab:hardware-plans} indicate stable execution, supporting the reliability of measured times and \sysname{}'s analytical model. As shown in Fig.~\ref{fig:cummulative-cost}, even modest per-iteration improvements can yield substantial savings over thousands of training iterations, reducing total training time by days or weeks.

%
\subsection{Scalability}
\label{sec:eval:scalability}
Moving beyond real hardware, we study \textbf{RQ1}, \textbf{RQ2} through large-scale simulation, evaluating how \sysname{}'s advantage over impr. Metis varies with two problem dimensions: the number of available GPUs (Table~\ref{tab:scalability_gpu_count}) and model depth (Table~\ref{tab:scalability_model_depth}). In each experiment, all other factors are held constant, so any change in the performance gap can be attributed solely to the parameter being varied. All reported times are iteration times based on the cost model.
Gain\,\% is $(T_{\text{Metis}} - T_{\text{MA}})/T_{\text{Metis}}$; positive values indicate that \sysname{} is faster. Across all ten configurations, \sysname{} matches or outperforms the improved state-of-the-art baseline.


\paragraphbe{GPU count.} The advantage over impr. Metis is non-monotonic with GPU pool size but becomes pronounced beyond $N=8$ (Table~\ref{tab:scalability_gpu_count}). For $N\leq6$, both perform similarly (gap $\leq0.2\%$), as limited heterogeneity leaves little optimization opportunity. For $N\geq8$, \sysname{} outperforms impr. Metis by $6.8$-$9.4\%$ through guided search that excludes slow devices and rebalances stages, while impr. Metis uses random subset sampling and proportional partitioning that can retain stragglers.


\begin{table}[t]
  \centering
  \caption{Iteration time vs.~GPU count ($L=12$, $H=2048$, fixed
    heterogeneity mix). 
    Positive favours \sysname{}.}
  \label{tab:scalability_gpu_count}
  \small
  \resizebox{0.75\columnwidth}{!}{
  \setlength{\tabcolsep}{5pt}
  \begin{tabular}{l l r r r c r}
    \toprule
    & & \textbf{impr. Metis} & \multicolumn{3}{c}{\textbf{\sysname{}}} & \textbf{Gain} \\
    \cmidrule(lr){4-6}
    \textbf{GPUs} & \textbf{Layers} & (ms) & Best & Mean & CI & (\%) \\
    \midrule
     4 & 12 & 1979.8 & 1979.8 & 1979.8 & $\pm$0.0 & +0.0 \\
     6 & 12 & 1754.2 & 1750.2 & 1750.2 & $\pm$0.0 & +0.2 \\
     8 & 12 & 1354.5 & 1254.8 & 1254.8 & $\pm$0.0 & +7.4 \\
    10 & 12 & 1323.6 & 1233.4 & 1233.4 & $\pm$0.0 & +6.8 \\
    12 & 12 & 1099.5 &  996.6 &  996.6 & $\pm$0.0 & +9.4 \\
    \bottomrule
  \end{tabular}
  }
\vspace{-1mm}
\end{table}

\begin{table}[t]
  \centering
  \caption{Iteration time vs.~model depth (GPUs\,$=12$, $H=2048$, fixed
    heterogeneity mix). 
    Positive favours \sysname{}.}
  \label{tab:scalability_model_depth}
  \small
  \resizebox{0.75\columnwidth}{!}{
  \setlength{\tabcolsep}{5pt}
  \begin{tabular}{l l r r r c r}
    \toprule
    & & \textbf{impr. Metis} & \multicolumn{3}{c}{\textbf{\sysname{}}} & \textbf{Gain} \\
    \cmidrule(lr){4-6}
    \textbf{GPUs} & \textbf{L} & (ms) & Best & Mean & CI & (\%) \\
    \midrule
    12 &  8 &  807.4 &  801.9 &  801.9 & $\pm$0.0  & +0.7  \\
    12 & 12 & 1099.5 &  996.6 &  996.6 & $\pm$0.0  & +9.4  \\
    12 & 24 & 1822.1 & 1705.9 & 1705.9 & $\pm$0.0  & +6.4  \\
    12 & 32 & 2204.8 & 2199.3 & 2199.3 & $\pm$0.0  & +0.3  \\
    12 & 48 & 2191.8 & 1958.7 & 1998.6 & $\pm$38.7 & +10.6 \\
    \bottomrule
  \end{tabular}
  }
\end{table}
\paragraphbe{Model depth.}
Model depth alone does not determine the performance gap
(Table~\ref{tab:scalability_model_depth}); the gain depends on how well the FLOPS-proportional Metis baseline balances the heterogeneous pipeline. When the proportional split is already near-balanced ($L=8,32$), the gap remains below $1\%$. Otherwise, proportional partitioning creates a straggler stage that \sysname{} improves through topology-aware exploration and rebalancing, yielding $6.4$--$10.6\%$ gains.

\paragraphbe{Convergence stability.}
Across all but the largest configuration ($L=48$), \sysname{} converges to the same solution across five seeds, indicating stable convergence. Variance appears only at $L=48$ ($\pm38.7$ms), where the larger search space introduces multiple local optima. Nevertheless, \sysname{} consistently outperforms the baseline on average, demonstrating robustness under increased optimization complexity.
\subsection{Optimality and Tractability}
\label{sec:eval:exact}

While Sect.~\ref{sec:eval:scalability} evaluates \sysname{}
against impr. Metis, \textbf{RQ1} and \textbf{RQ3} require comparison with the global optimum. We therefore use two exact solvers as ground truth: Brute-Force (BF) and Branch-and-Bound (B\&B).

\paragraphbe{Correctness on heterogeneous scenarios.}
We first evaluate \sysname{} on four simulated heterogeneous scenarios (3 to 6 GPUs, four GPU types, $L=12$). As Table~\ref{tab:exact-correctness} shows, \sysname{} matches the exact optimum on all four scenarios across all ten seeds, yielding a 100.00\% optimality (Opt.).

\begin{table}[t]
  \centering
  \caption{\sysname{} recovers the exact optimum on every computable
    scenario. 
    Match counts seeds reaching optimum.}
  \label{tab:exact-correctness}
  \small
  \resizebox{0.75\columnwidth}{!}{
  \setlength{\tabcolsep}{6pt}
  \begin{tabular}{c c r r c c}
    \toprule
    \textbf{N} & \textbf{L} & \textbf{$T^{*}$ (ms)}
      & \textbf{\sysname{} (ms)} & \textbf{Optimality} & \textbf{Match} \\
    \midrule
    3 & 12 & 5291.165 & 5291.165 & \textbf{100\%} & 10/10 \\
    4 & 12 & 1979.826 & 1979.826 & \textbf{100\%} & 10/10 \\
    5 & 12 & 1878.830 & 1878.830 & \textbf{100\%} & 10/10 \\
    6 & 12 & 1750.161 & 1750.161 & \textbf{100\%} & 10/10 \\
    \bottomrule
  \end{tabular}
  }
  \vspace{-2mm}
\end{table}

\begin{figure*}[t]
  \centering
  \begin{subfigure}[b]{0.33\textwidth}
    \centering
    \includegraphics[width=\linewidth]{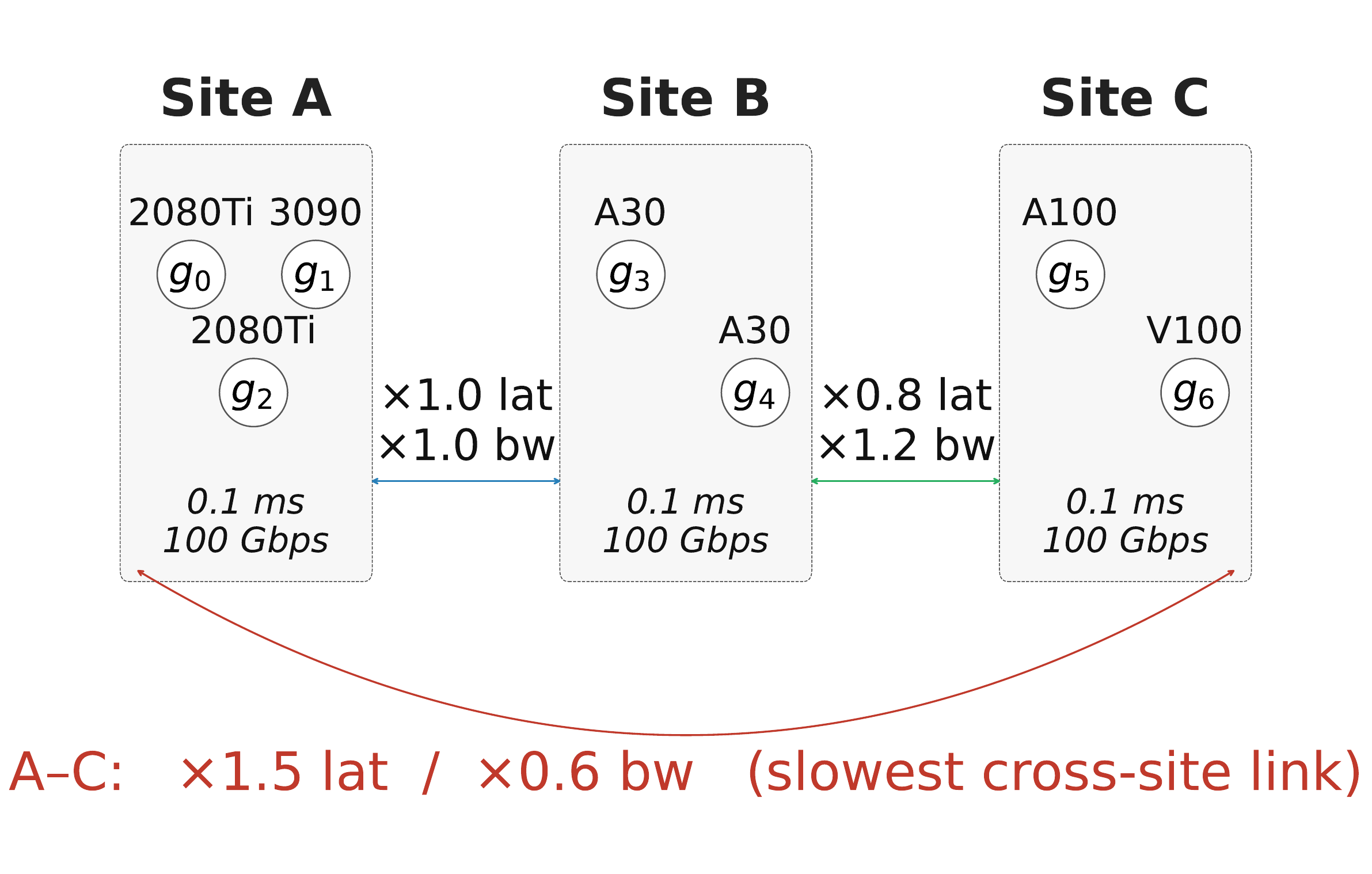}
    \caption{Three-site topology}
    \label{fig:net-topo}
  \end{subfigure}
  \hfill
  \begin{subfigure}[b]{0.3\textwidth}
    \centering
    \includegraphics[width=\linewidth]{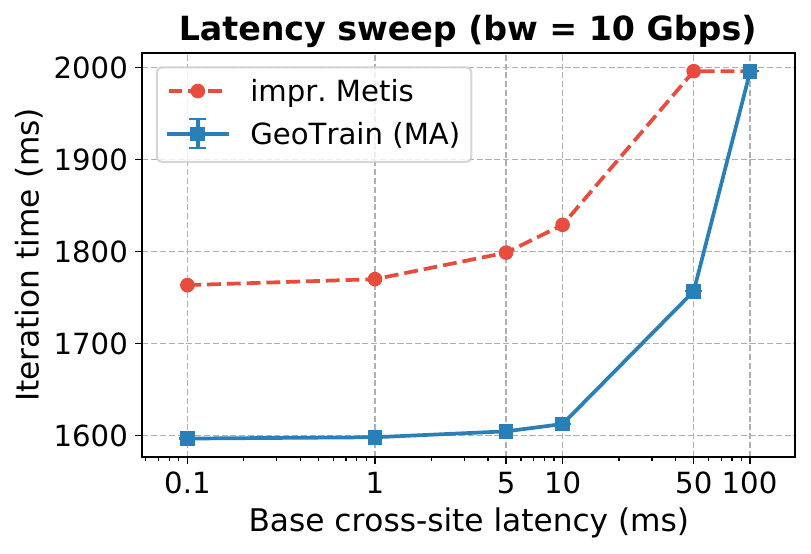}
    \caption{Latency sweep (bw\,=\,10\,Gbps)}
    \label{fig:net-latency}
  \end{subfigure}
  \hfill
  \begin{subfigure}[b]{0.3\textwidth}
    \centering
    \includegraphics[width=\linewidth]{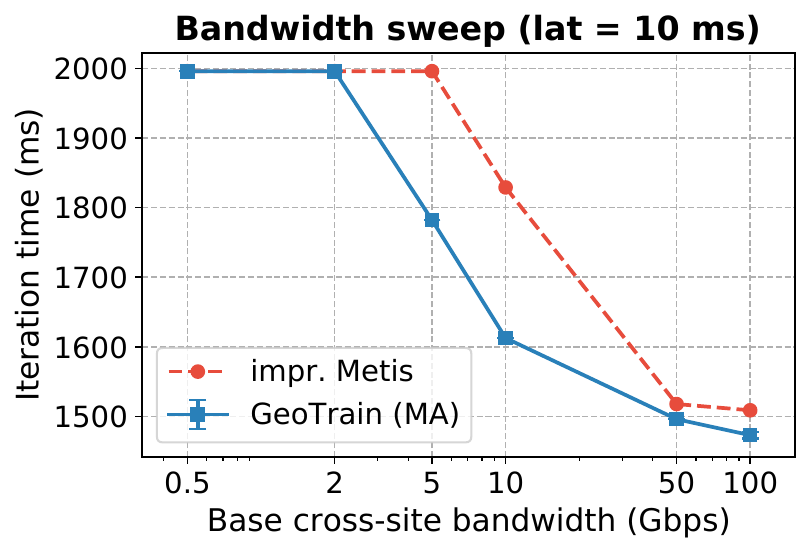}
    \caption{Bandwidth sweep (lat\,=\,10\,ms)}
    \label{fig:net-bandwidth}
  \end{subfigure}
  \caption{Network sensitivity}
  \label{fig:network}
  \vspace{-3mm}
\end{figure*}

\paragraphbe{Tractability and optimality}
We vary GPU count ($N=2$-$12$, $L=12$) and compare BF, B\&B, and \sysname{} (Table~\ref{tab:exact-tractability}). The search space grows rapidly with $N$, from just $24$ configurations at $N=2$ to over $40$\,billion at $N=12$. where $f(N,L)=\sum_{k=1}^{\min(N,L)} \binom{L-1}{k-1}\,P(N,k)$, counts all valid layer partitions and GPU assignments. 
On all tractable instances ($N\in\{2,4,6\}$), \sysname{} matches the optimum, yielding a 100\% optimality. Beyond $N=6$, BF and B\&B fail to finish within the 120\,s budget, while \sysname{} remains the only method that returns a plan.

\begin{table}[t]
  \centering
  \caption{Iteration time and planning time and  vs.\ GPU count ($L=12$).
    Exact solvers exhaust the 120\,s budget at $N\!\geq\!8$ (---).}
  \label{tab:exact-tractability}
  \small
 \resizebox{.95\columnwidth}{!}{
  \setlength{\tabcolsep}{5pt}
  \begin{tabular}{c c r r r r r r r}
    \toprule
    & & & \multicolumn{2}{c}{\textbf{BF}}
        & \multicolumn{2}{c}{\textbf{B\&B}}
        & \multicolumn{2}{c}{\textbf{\sysname{}}} \\
    \cmidrule(lr){4-5}\cmidrule(lr){6-7}\cmidrule(lr){8-9}
    \textbf{N} & \textbf{L} & \textbf{Space}
        & iter. (ms) & plan. & iter. (ms) & plan. & iter. (ms) & plan. \\
    \midrule
    2  & 12 & 24      & 5839.7 & 0.0\,s  & 5839.7 & 0.0\,s  & 5839.7 & 13.3\,s \\
    4  & 12 & 5.4K    & 1979.8 & 0.4\,s  & 1979.8 & 0.3\,s  & 1979.8 & 15.9\,s \\
    6  & 12 & 637K    & 1750.2 & 16.6\,s & 1750.2 & 15.9\,s & 1750.2 & 18.8\,s \\
    8  & 12 & 43.8M   & ---    & ---     & ---    & ---     & 1254.8 & 22.8\,s \\
    10 & 12 & 1.8B    & ---    & ---     & ---    & ---     & 1233.4 & 32.5\,s \\
    12 & 12 & 40.9B   & ---    & ---     & ---    & ---     &  996.6 & 29.4\,s \\
    \bottomrule
  \end{tabular}}
\vspace{-3mm}
\end{table}


Together, these results demonstrate \sysname{}'s quality: it matches the verifiable optimum on tractable instances and remains effective beyond the reach of exact methods.

\subsection{Network Sensitivity}
\label{sec:eval:network}

We evaluate how planners adapt to geo-distributed networks and when topology-aware layer allocation yields the largest gains over the baseline (\textbf{RQ4}). Using a simulated three-site topology (Fig.~\ref{fig:net-topo}), we isolate cross-site effects: intra-site links provide 0.1,ms latency and 100,Gbps bandwidth, while cross-site links are asymmetric: A-B at $1.0\times$, B-C at $0.8\times$ latency/$1.2\times$ bandwidth, and A-C at $1.5\times$ latency/$0.6\times$ bandwidth. Base latency and bandwidth vary from 0.1-100,ms and 0.5-100,Gbps, respectively. We evaluate \sysname{} and impr. Metis across these settings, averaging 10 runs.



\paragraphbe{Latency sensitivity.} We fix cross-site bandwidth at 10\,Gbps and sweep the base cross-site latency
(Fig.~\ref{fig:net-latency}). Both planners exclude Site~A at every level: its GPUs are weakest in compute and carry the highest cross-site penalty. From 0.1\,ms to 10\,ms, both select multi-site B+C plans with a single cross-site hop, and \sysname{} leads impr. Metis by 10.4\% to 13.4\% via better load balance.

\begin{figure}[t]
  \centering
  \includegraphics[width=1\linewidth]{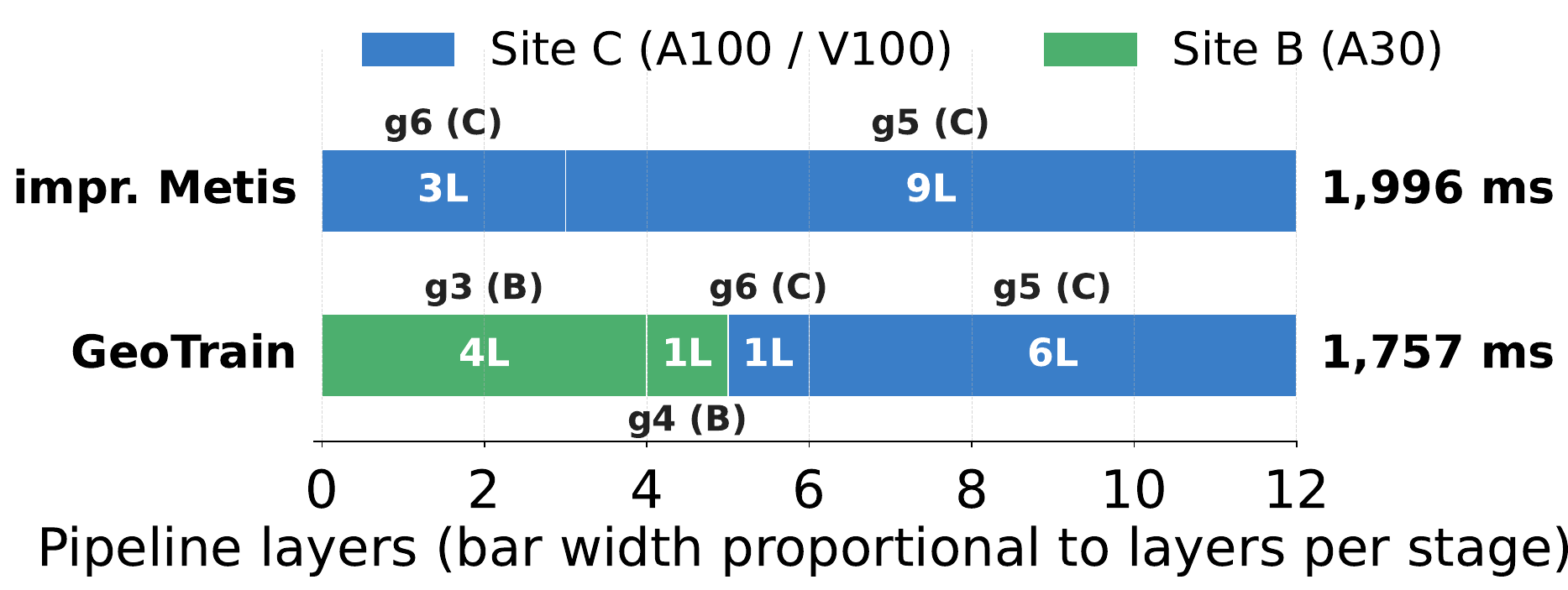}
  \caption{Plans found at 50\,ms cross-site latency.}
  \vspace{-2mm}
  \label{fig:net-plan}
\end{figure}

At the 50\,ms peak (13.6\%), impr. Metis abandons cross-site communication for a single-site plan on Site~C (Fig.~\ref{fig:net-plan}, top), as its proportional rule cannot justify a multi-site split. \sysname{} instead retains a balanced B+C plan (Fig.~\ref{fig:net-plan}, bottom), where load balance offsets communication cost, cutting iteration time 1{,}996 to 1{,}757\,ms. This highlights the value of topology-aware layer allocation. At 100\,ms, both converge to Site~C and the gap closes to 0\%.

\paragraphbe{Bandwidth sensitivity.}
We fix cross-site latency at 10\,ms and sweep bandwidth from $0.5$ to $100$\,Gbps (Fig.~\ref{fig:net-bandwidth}). At low bandwidth, communication dominates and both planners consolidate onto a single site. As bandwidth increases, multi-site execution becomes viable and \sysname{} exploits it faster than impr. Metis, opening its largest advantage ($12.0\%$ at 5\,Gbps). Once bandwidth is abundant, communication is no longer a bottleneck, and the two converge.

Overall, \sysname{} tracks network conditions more closely, shifting between single- and multi-site plans at finer granularity to capture gains missed by the baselines.
\subsection{Energy-Time Trade-off}
\label{sec:eval:pareto}
\begin{figure}[t]
  \centering
  \includegraphics[width=0.7\columnwidth, height=4.5cm]{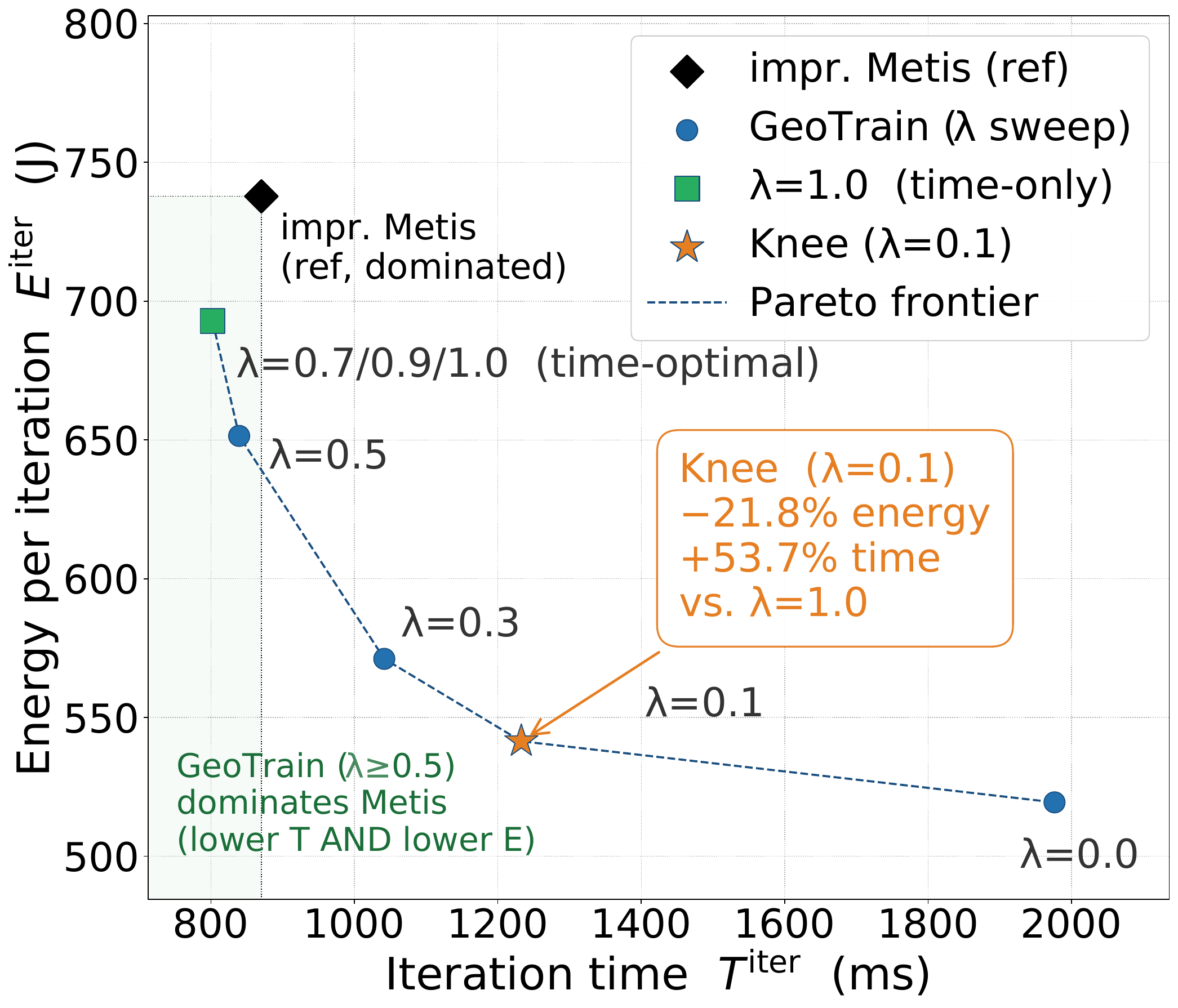}
  \caption{%
    Energy-time Pareto frontier of \sysname{} (14 GPUs, 12 layers). Points show the best plan across five seeds for each $\lambda$. Dashed line
    connects the frontier.}
    \vspace{-3mm}
  \label{fig:pareto}
\end{figure}

\begin{figure}[t]
  \centering
  \includegraphics[width=\columnwidth, height=4cm]{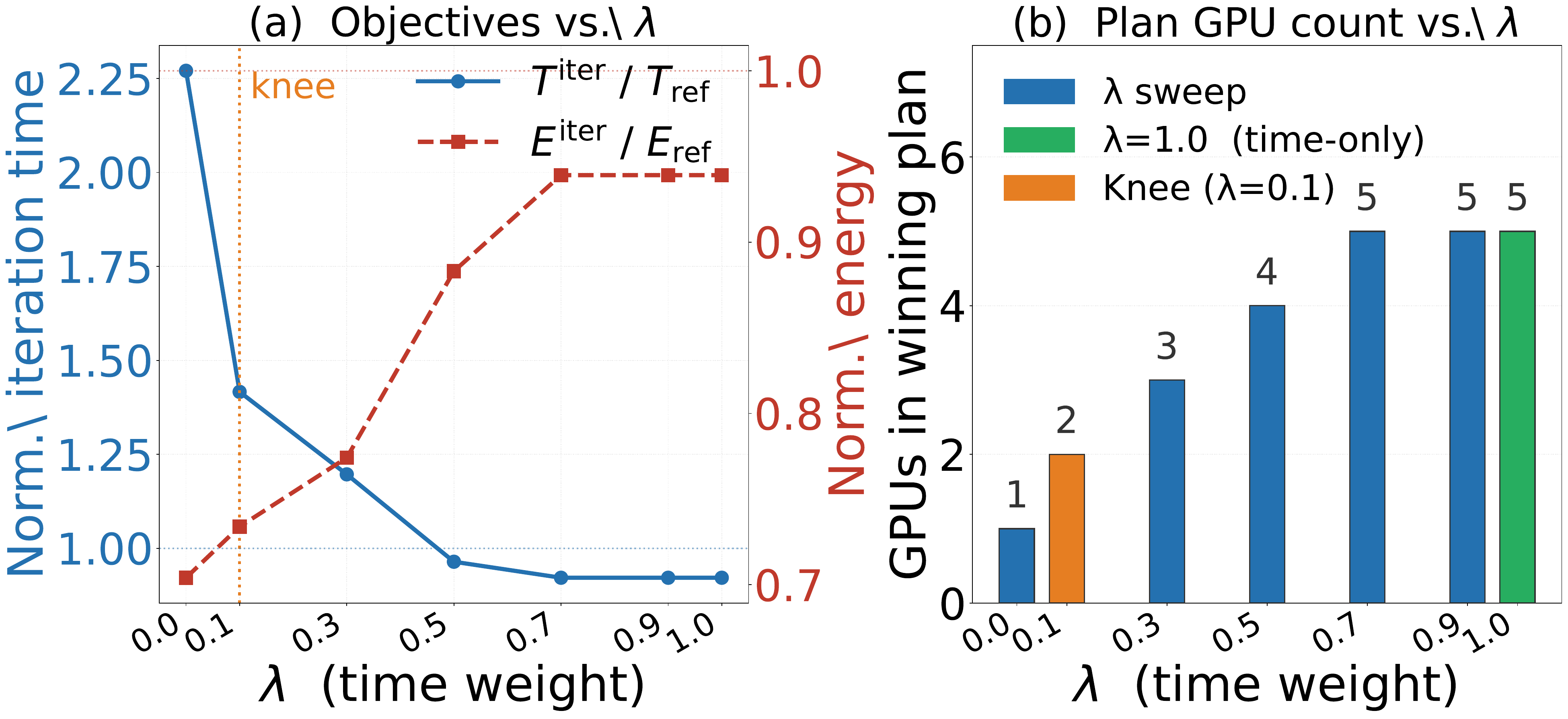}
  \caption{%
    \textbf{(a)}~Normalised iteration time ($T^{\mathrm{iter}}/T_{\mathrm{ref}}$,
    blue) and energy ($E^{\mathrm{iter}}/E_{\mathrm{ref}}$, red) vs.\ $\lambda$;
    dotted lines mark the impr. Metis reference ($=1$).
    \textbf{(b)}~GPUs in the winning plan vs.\ $\lambda$.}
    \vspace{-3mm}
  \label{fig:pareto-summary}
\end{figure}

The objective in Eq.~\ref{eq:objective} enables \sysname{} to explore the Pareto frontier between iteration time and energy via $\lambda \in [0,1]$, from pure energy ($\lambda\!=\!0$) to pure time minimization ($\lambda\!=\!1$). To answer \textbf{RQ5}, we vary $\lambda$ on a 14-GPU, 12-layer simulated setup, with five seeds per value. The GPU types are A30 (24~GB), A100~PCIe (40~GB), and A100~SXM (80~GB), with idle/active powers of 25/165~W, 40/250~W, and 60/400~W, respectively. We use rated TDP 
as active power and set idle power to approximately 15\% of TDP, consistent with the $\sim$60~W idle power measured for the A100~SXM (80~GB)~\cite{nvidia-a30-datasheet, nvidia-a100-pcie-brief, nvidia-a100-datasheet, hu2024characterization}.
As the objectives differ in units, we normalize them using $T_{\mathrm{ref}}=870.4$\,ms and $E_{\mathrm{ref}}=737.8$\,J from the impr. Metis plan. Table~\ref{tab:pareto} reports the best $(T^{\mathrm{iter}}, E^{\mathrm{iter}})$ per $\lambda$, while Fig.~\ref{fig:pareto} shows the frontier and Fig.~\ref{fig:pareto-summary} shows how objectives and GPU subset size vary with $\lambda$. All seeds converge to the same plan.

\begin{table}[t]
  \centering
  \caption{%
    Energy-time Pareto frontier (14 GPUs, 12 layers, $H\!=\!2048$).
    $\Delta T$ and $\Delta E$ are deviations from the $\lambda\!=\!1.0$ plan.}
  \label{tab:pareto}
  \small
  \setlength{\tabcolsep}{4.5pt}
  \resizebox{0.95\columnwidth}{!}{%
  \begin{tabular}{c r r r r l}
    \toprule
    $\boldsymbol{\lambda}$
      & $\boldsymbol{T^{\mathrm{iter}}}$ \textbf{(ms)}
      & $\boldsymbol{E^{\mathrm{iter}}}$ \textbf{(J)}
      & $\boldsymbol{\Delta T}$
      & $\boldsymbol{\Delta E}$
      & \textbf{GPUs selected} \\
    \midrule
    0.0
      & 1{,}976.2 & 519.4
      & $+$146.3\% & $-$25.0\%
      & 1$\times$A100\,PCIe \\
    0.1$^{\star}$
      & 1{,}232.7 & 541.4
      & $+$53.7\%  & $-$21.8\%
      & 2$\times$A100\,PCIe \\
    0.3
      & 1{,}041.7 & 571.1
      & $+$29.8\%  & $-$17.6\%
      & 3$\times$A100\,PCIe \\
    0.5
      &   839.3   & 651.5
      & $+$4.6\%   & $-$6.0\%
      & 2$\times$A100\,SXM, 2$\times$A100\,PCIe \\
    0.7
      &   802.3   & 692.8
      & ---        & ---
      & 2$\times$A100\,SXM, 3$\times$A100\,PCIe \\
    0.9
      &   802.3   & 692.8
      & ---        & ---
      & 2$\times$A100\,SXM, 3$\times$A100\,PCIe \\
    1.0
      &   802.3   & 692.8
      & \emph{(baseline)} & \emph{(baseline)}
      & 2$\times$A100\,SXM, 3$\times$A100\,PCIe \\
    \midrule
    \multicolumn{6}{l}{\itshape impr. Metis (reference):} \\
    ---
      & 870.4  & 737.8
      & $+$8.5\%   & $+$6.5\%
      & 3$\times$A100\,SXM, 2$\times$A100\,PCIe \\
    \bottomrule
    \multicolumn{6}{l}{%
      \footnotesize
      $^{\star}$~Knee of the frontier (maximum perpendicular distance from the
      line joining the extreme points).}
  \end{tabular}
  }
\end{table}

\paragraphbe{Energy-saving ($\lambda \leq 0.3$).} \sysname{} uses as few GPUs as possible, selecting A100\,PCIe devices to minimise idle power from unused hardware. It notably avoids A30 GPUs despite their lower TDP, because A100\,PCIe achieves better compute efficiency-lower peak power per effective FLOPS - so faster stage execution reduces total
active-phase energy

\paragraphbe{Balanced ($\lambda = 0.5$).} \sysname{} augments the A100\,PCIe
stages with two A100\,SXM GPUs (4-GPU plan). Compared to the time-optimal plan ($\lambda\!=\!1$), it is only 4.6\% slower but uses 6.0\% less energy, making it a strong energy-aware operating point with near-optimal speed.

\paragraphbe{Time-optimal ($\lambda \geq 0.7$).} All five seeds converge to an
identical 5-GPU plan for $\lambda \in \left[ 0.7, 1.0\right] $, indicating that the time-optimal configuration is a stable fixed point and that
increasing $\lambda$ beyond 0.7 does not change the solution.


\paragraphbe{\sysname{} strictly dominates Metis.}
The impr. Metis plan (870.4,ms, 737.8,J) is dominated by all \sysname{} plans with $\lambda \geq 0.5$. At $\lambda=0.5$, \sysname{} reduces iteration time and energy by $3.6\%$ and $11.7\%$, respectively. At $\lambda=1.0$, it achieves the fastest configuration (802.3,ms) while still reducing energy by $6.1\%$.


\paragraphbe{The knee gives the best return on energy.}
The maximum-perpendicular-distance criterion~\cite{satopaa2011kneedle}
places the knee at $\lambda\!=\!0.1$, whose two-GPU plan cuts iteration time by 37.6\% over the single-GPU energy-only
plan ($\lambda\!=\!0$) for only 4.2\% more energy. 
It is the recommended point for energy-constrained deployments needing high throughput. Those requiring near-optimal speed can use $\lambda\!=\!0.5$, within 4.6\% of the time-only optimum while saving 6.0\% energy.
\section{Conclusion and Future Work}
\label{sec:conclusion}
We presented GeoTrain, an automatic parallelism system for foundation model training on heterogeneous, geo-distributed GPUs. By jointly optimizing GPU selection, model partitioning, and placement under compute, memory, power, and network constraints, GeoTrain efficiently explores the combinatorial search space using an optimized memetic algorithm. Across real and simulated environments, GeoTrain outperforms state-of-the-art baselines, reducing iteration time and energy by up to $61.4\%$ and $11.7\%$,
respectively, while matching the verifiable optimum on small instances and scaling beyond exact methods. These results show that topology- and energy-aware planning can provide a practical and scalable approach to distributed foundation model training.
In future work, we will extend GeoTrain to support additional parallelism techniques within pipeline stages, adapt to dynamic network and resource conditions, incorporate additional practical objectives, and investigate larger-scale production deployments.

\bibliographystyle{ACM-Reference-Format}
\bibliography{references}

@inproceedings{brown2020language,
  author    = {Tom Brown and Benjamin Mann and Nick Ryder and others},
  title     = {Language Models are Few-Shot Learners},
  booktitle = {Advances in Neural Information Processing Systems (NeurIPS)},
  volume    = {33},
  pages     = {1877--1901},
  year      = {2020},
}

@misc{touvron2023llama,
  author = {Hugo Touvron and Thibaut Lavril and Gautier Izacard and others},
  title  = {LLaMA: Open and Efficient Foundation Language Models},
  year   = {2023},
  note   = {arXiv preprint arXiv:2302.13971},
}

@inproceedings{zheng2022alpa,
  author    = {Lianmin Zheng and Zhuohan Li and Hao Zhang and others},
  title     = {Alpa: Automating Inter- and Intra-Operator Parallelism for Distributed Deep Learning},
  booktitle = {16th USENIX Symposium on Operating Systems Design and Implementation (OSDI 22)},
  pages     = {559--578},
  year      = {2022},
}

@article{li2022amp,
  author  = {Dacheng Li and Hongyi Wang and Eric Xing and Hao Zhang},
  title   = {AMP: Automatically Finding Model Parallel Strategies with Heterogeneity Awareness},
  journal = {Advances in Neural Information Processing Systems (NeurIPS)},
  volume  = {35},
  pages   = {6630--6639},
  year    = {2022},
}

@inproceedings{um2024metis,
  author    = {Taegeon Um and Byungsoo Oh and Minyoung Kang and others},
  title     = {Metis: Fast Automatic Distributed Training on Heterogeneous GPUs},
  booktitle = {2024 USENIX Annual Technical Conference (USENIX ATC 24)},
  pages     = {563--578},
  year      = {2024},
}

@article{miao2022galvatron,
author = {Miao, Xupeng and Wang, Yujie and Jiang, Youhe and Shi, Chunan and Nie, Xiaonan and Zhang, Hailin and Cui, Bin},
title = {Galvatron: Efficient Transformer Training over Multiple GPUs Using Automatic Parallelism},
year = {2022},
issue_date = {November 2022},
publisher = {VLDB Endowment},
volume = {16},
number = {3},
issn = {2150-8097},
url = {https://doi.org/10.14778/3570690.3570697},
doi = {10.14778/3570690.3570697},
journal = {Proc. VLDB Endow.},
month = nov,
pages = {470–479},
numpages = {10}
}

@inproceedings{shoeybi2019megatron,
  author = {Mohammad Shoeybi and Mostofa Patwary and Raul Puri and others},
  title  = {Megatron-LM: Training Multi-Billion Parameter Language Models Using Model Parallelism},
  year   = {2019},
  note   = {arXiv preprint arXiv:1909.08053},
}

@inproceedings{park2020hetpipe,
  author    = {Jay H. Park and Gyeongchan Yun and Chang M. Yi and others},
  title     = {HetPipe: Enabling Large DNN Training on (Whimpy) Heterogeneous GPU Clusters through Integration of Pipelined Model Parallelism and Data Parallelism},
  booktitle = {2020 USENIX Annual Technical Conference (USENIX ATC 20)},
  pages     = {307--321},
  year      = {2020},
}

@inproceedings{narayanan2019pipedream,
  author    = {Deepak Narayanan and Aaron Harlap and Amar Phanishayee and others},
  title     = {PipeDream: Generalized Pipeline Parallelism for DNN Training},
  booktitle = {Proceedings of the 27th ACM Symposium on Operating Systems Principles (SOSP 19)},
  pages     = {1--15},
  year      = {2019},
}

@inproceedings{huang2019gpipe,
  author    = {Yanping Huang and Youlong Cheng and Ankur Bapna and others},
  title     = {GPipe: Efficient Training of Giant Neural Networks using Pipeline Parallelism},
  booktitle = {Advances in Neural Information Processing Systems (NeurIPS)},
  volume    = {32},
  year      = {2019},
}

@inproceedings{strubell2019energy,
  author    = {Emma Strubell and Ananya Ganesh and Andrew McCallum},
  title     = {Energy and Policy Considerations for Deep Learning in NLP},
  booktitle = {Proceedings of the 57th Annual Meeting of the Association for Computational Linguistics (ACL)},
  pages     = {3645--3650},
  year      = {2019},
}

@misc{patterson2021carbon,
  author = {David Patterson and Joseph Gonzalez and Quoc Le and others},
  title  = {Carbon Emissions and Large Neural Network Training},
  year   = {2021},
  note   = {arXiv preprint arXiv:2104.10350},
}

@inproceedings{you2023zeus,
  author    = {Jie You and Jae-Won Chung and Mosharaf Chowdhury},
  title     = {Zeus: Understanding and Optimizing GPU Energy Consumption of DNN Training},
  booktitle = {20th USENIX Symposium on Networked Systems Design and Implementation (NSDI 23)},
  pages     = {119--139},
  year      = {2023},
}

@inproceedings{rajbhandari2020zero,
  author    = {Samyam Rajbhandari and Jeff Rasley and Olatunji Ruwase and Yuxiong He},
  title     = {{ZeRO}: Memory Optimizations Toward Training Trillion Parameter Models},
  booktitle = {Proceedings of the International Conference for High Performance Computing,
               Networking, Storage and Analysis (SC 20)},
  year      = {2020},
}

@article{kaplan2020scaling,
  author  = {Jared Kaplan and Sam McCandlish and Tom Henighan and others},
  title   = {Scaling Laws for Neural Language Models},
  journal = {arXiv preprint arXiv:2001.08361},
  year    = {2020},
}

@inproceedings{yuan2022dtfm,
  author    = {Yuan, Binhang and He, Yongjun and Davis, Jared Quincy and Zhang, Tianyi and Dao, Tri and Chen, Beidi and Liang, Percy and Re, Christopher and Zhang, Ce},
  title     = {Decentralized Training of Foundation Models in Heterogeneous Environments},
  booktitle = {Proceedings of the 36th International Conference on Neural Information Processing Systems (NeurIPS '22)},
  year      = {2022},
  publisher = {Curran Associates Inc.},
  address   = {New Orleans, LA, USA},
  isbn      = {9781713871088}
}

@inproceedings{ryabinin2023swarm,
  author    = {Ryabinin, Max and Dettmers, Tim and Diskin, Michael and Borzunov, Alexander},
  title     = {{SWARM} Parallelism: Training Large Models Can Be Surprisingly Communication-Efficient},
  booktitle = {Proceedings of the 40th International Conference on Machine Learning (ICML '23)},
  year      = {2023},
  series    = {ICML'23},
  publisher = {JMLR.org},
  address   = {Honolulu, Hawaii, USA}
}

@inproceedings{strati2025sailor,
author    = {Strati, Foteini and Zhang, Zhendong and Manos, George and
               S{\'a}nchez P{\'e}riz, Ixeia and Hu, Qinghao and Chen, Tiancheng and
               Buzcu, Berk and Han, Song and Delgado, Pamela and Klimovic, Ana},
  title     = {Sailor: Automating Distributed Training over Dynamic,
               Heterogeneous, and Geo-distributed Clusters},
  booktitle = {Proceedings of the ACM SIGOPS 31st Symposium on Operating
               Systems Principles (SOSP '25)},
  pages     = {204--220},
  publisher = {Association for Computing Machinery},
  address   = {Seoul, Republic of Korea},
  year      = {2025},
  doi       = {10.1145/3731569.3764839}
}

@inproceedings{wang2023cocktailsgd,
  author    = {Wang, Jue and Lu, Yucheng and Yuan, Binhang and Chen, Beidi and Liang, Percy and De Sa, Christopher and Re, Christopher and Zhang, Ce},
  title     = {{CocktailSGD}: Fine-tuning Foundation Models over 500{Mbps} Networks},
  booktitle = {Proceedings of the 40th International Conference on Machine Learning (ICML '23)},
  year      = {2023},
  series    = {ICML'23},
  publisher = {JMLR.org},
  address   = {Honolulu, Hawaii, USA}
}

@inproceedings{vaswani2017attention,
  author    = {Vaswani, Ashish and Shazeer, Noam and Parmar, Niki and Uszkoreit, Jakob and Jones, Llion and Gomez, Aidan N. and Kaiser, {\L}ukasz and Polosukhin, Illia},
  title     = {Attention Is All You Need},
  booktitle = {Advances in Neural Information Processing Systems (NeurIPS)},
  volume    = {30},
  year      = {2017},
  pages     = {5998--6008}
}

@inproceedings{narayanan2021megatron,
  author    = {Narayanan, Deepak and Shoeybi, Mohammad and Casper, Jared and LeGresley, Patrick and Patwary, Mostofa and Korthikanti, Vijay Anand and Vainbrand, Dmitri and Kashinkunti, Prethvi and Bernauer, Julie and Catanzaro, Bryan and Phanishayee, Amar and Zaharia, Matei},
  title     = {Efficient Large-Scale Language Model Training on {GPU} Clusters Using {Megatron-LM}},
  booktitle = {Proceedings of the International Conference for High Performance Computing, Networking, Storage and Analysis (SC '21)},
  year      = {2021},
  pages     = {1--14},
  doi       = {10.1145/3458817.3476209}
}

@inproceedings{kingma2015adam,
  author    = {Kingma, Diederik P. and Ba, Jimmy},
  title     = {{Adam}: A Method for Stochastic Optimization},
  booktitle = {Proceedings of the 3rd International Conference on Learning Representations (ICLR)},
  year      = {2015}
}

@incollection{moscato2003memetic,
  author    = {Moscato, Pablo and Cotta, Carlos},
  title     = {A Gentle Introduction to Memetic Algorithms},
  booktitle = {Handbook of Metaheuristics},
  editor    = {Glover, Fred and Kochenberger, Gary A.},
  series    = {International Series in Operations Research \& Management Science},
  volume    = {57},
  pages     = {105--144},
  publisher = {Springer},
  address   = {Boston, MA},
  year      = {2003},
  doi       = {10.1007/0-306-48056-5_5}
}

@inproceedings{satopaa2011kneedle,
  author    = {Satop{\"a}{\"a}, Ville and Albrecht, Jeannie and Irwin, David and Raghavan, Barath},
  title     = {Finding a ``Kneedle'' in a Haystack: Detecting Knee Points in System Behavior},
  booktitle = {2011 31st International Conference on Distributed Computing Systems Workshops (ICDCSW)},
  year      = {2011},
  pages     = {166--171},
  publisher = {IEEE},
  doi       = {10.1109/ICDCSW.2011.20}
}

@article{miller1995tournament,
  author  = {Miller, Brad L. and Goldberg, David E.},
  title   = {Genetic Algorithms, Tournament Selection, and the Effects of Noise},
  journal = {Complex Systems},
  volume  = {9},
  number  = {3},
  pages   = {193--212},
  year    = {1995}
}

@incollection{syswerda1991steadystate,
  author    = {Syswerda, Gilbert},
  title     = {A Study of Reproduction in Generational and Steady-State Genetic Algorithms},
  booktitle = {Foundations of Genetic Algorithms},
  editor    = {Rawlins, Gregory J. E.},
  volume    = {1},
  pages     = {94--101},
  publisher = {Morgan Kaufmann},
  year      = {1991},
  doi       = {10.1016/B978-0-08-050684-5.50009-4}
}

@article{zong2025hgtrainer,
  author  = {Zong, Zan and Guo, Minkun and Zhai, Mingshu and Tang, Yinan and Li, Jianjiang and Zhai, Jidong},
  title   = {Training Large Models on Heterogeneous and Geo-Distributed Resource with Constricted Networks},
  journal = {Big Data Mining and Analytics},
  volume  = {8},
  number  = {4},
  pages   = {966--980},
  year    = {2025},
  doi     = {10.26599/BDMA.2025.9020031}
}

@inproceedings{jang2023oobleck,
  author    = {Insu Jang and Zhenning Yang and Zhen Zhang and Xin Jin and Mosharaf Chowdhury},
  title     = {Oobleck: Resilient Distributed Training of Large Models Using Pipeline Templates},
  booktitle = {Proceedings of the 29th Symposium on Operating Systems Principles (SOSP '23)},
  year      = {2023},
  pages     = {382--395},
  publisher = {Association for Computing Machinery},
  doi       = {10.1145/3600006.3613152}
}

@inproceedings{sun2024adapipe,
  author    = {Zhenbo Sun and Huanqi Cao and Yuanwei Wang and Guanyu Feng and Shengqi Chen and Haojie Wang and Wenguang Chen},
  title     = {{AdaPipe}: Optimizing Pipeline Parallelism with Adaptive Recomputation and Partitioning},
  booktitle = {Proceedings of the 29th ACM International Conference on Architectural Support for Programming Languages and Operating Systems (ASPLOS '24), Volume 3},
  year      = {2024},
  publisher = {Association for Computing Machinery},
  doi       = {10.1145/3620666.3651359}
}

@article{li2020pytorchddp,
  author    = {Shen Li and Yanli Zhao and Rohan Varma and Omkar Salpekar and Pieter Noordhuis and Teng Li and Adam Paszke and Jeff Smith and Brian Vaughan and Pritam Damania and Soumith Chintala},
  title     = {PyTorch Distributed: Experiences on Accelerating Data Parallel Training},
  journal   = {Proceedings of the VLDB Endowment},
  volume    = {13},
  number    = {12},
  pages     = {3005--3018},
  year      = {2020},
  doi       = {10.14778/3415478.3415530}
}

@article{blickle1996comparison,
  title={A Comparison of Selection Schemes Used in Evolutionary Algorithms},
  author={Blickle, Tobias and Thiele, Lothar},
  journal={Evolutionary Computation},
  volume={4},
  number={4},
  pages={361--394},
  year={1996},
  doi={10.1162/evco.1996.4.4.361}
}

@misc{nvidia-a100-datasheet,
  author       = {{NVIDIA Corporation}},
  title        = {{NVIDIA A100 Tensor Core GPU Datasheet}},
  year         = {2021},
  howpublished = {\url{https://www.nvidia.com/content/dam/en-zz/Solutions/Data-Center/a100/pdf/nvidia-a100-datasheet-nvidia-us-2188504-web.pdf}},
  note         = {Max Thermal Design Power: 250/300~W (PCIe), 400~W (SXM). Accessed 2026-06-11}
}

@misc{nvidia-a100-pcie-brief,
  author       = {{NVIDIA Corporation}},
  title        = {{NVIDIA A100 40GB PCIe GPU Accelerator Product Brief}},
  number       = {PB-10137-001_v03},
  year         = {2020},
  howpublished = {\url{https://www.nvidia.com/content/dam/en-zz/Solutions/Data-Center/a100/pdf/A100-PCIE-Prduct-Brief.pdf}},
  note         = {States the GPU operates unconstrained up to its TDP level of 250~W. Accessed 2026-06-11}
}

@misc{nvidia-a30-datasheet,
  author       = {{NVIDIA Corporation}},
  title        = {{NVIDIA A30 Tensor Core GPU Datasheet}},
  year         = {2021},
  howpublished = {\url{https://www.nvidia.com/content/dam/en-zz/Solutions/data-center/products/a30-gpu/pdf/a30-datasheet.pdf}},
  note         = {Max Thermal Design Power: 165~W. Accessed 2026-06-11}
}

@inproceedings{hu2024characterization,
  author    = {Qinghao Hu and Zhisheng Ye and Zerui Wang and Guoteng Wang
               and Meng Zhang and Qiaoling Chen and Peng Sun and Dahua Lin
               and Xiaolin Wang and Yingwei Luo and Yonggang Wen
               and Tianwei Zhang},
  title     = {Characterization of Large Language Model Development
               in the Datacenter},
  booktitle = {21st USENIX Symposium on Networked Systems Design
               and Implementation (NSDI 24)},
  pages     = {709--729},
  year      = {2024},
  address   = {Santa Clara, CA, USA},
  month     = apr,
  publisher = {USENIX Association},
  isbn      = {978-1-939133-39-7},
  url       = {https://www.usenix.org/conference/nsdi24/presentation/hu}
}

@inproceedings{jeon2025graphpipe,
author = {Jeon, Byungsoo and Wu, Mengdi and Cao, Shiyi and Kim, Sunghyun and Park, Sunghyun and Aggarwal, Neeraj and Unger, Colin and Arfeen, Daiyaan and Liao, Peiyuan and Miao, Xupeng and Alizadeh, Mohammad and Ganger, Gregory R. and Chen, Tianqi and Jia, Zhihao},
title = {GraphPipe: Improving Performance and Scalability of DNN Training with Graph Pipeline Parallelism},
year = {2025},
isbn = {9798400706981},
publisher = {Association for Computing Machinery},
address = {New York, NY, USA},
url = {https://doi.org/10.1145/3669940.3707220},
doi = {10.1145/3669940.3707220},
pages = {557–571},
numpages = {15},
location = {Rotterdam, Netherlands},
series = {ASPLOS '25}
}

\appendix
\appendix 
\setcounter{table}{0}
\section{Experimental Configurations}
\begin{center}
\captionof{table}{Transformer configuration details in real-world and simulated experiments}
\label{tab:configs}
\resizebox{0.8\columnwidth}{!}{%
\begin{tabular}{lcc}
\toprule
 & Real training & Simulated studies \\
\midrule
Hidden size $H$       & $1024$     & $2048$ \\
Depth $L$ (layers)    & $16$-$32$ & $8$-$48$ \\
Microbatch size $u$   & $16$       & $8$ \\
Microbatches $M$      & $64$       & $8$ \\
Batch size $B$        & $1024$     & $64$ \\
Sequence length $T$   & $128$      & $2048$ \\
Vocabulary size       & $50{,}258$ & $50{,}258$ \\
Precision             & FP32       & FP16 \\
\bottomrule
\end{tabular}%
}
\end{center}

\end{document}